\documentclass[a4paper,11pt,titlepage,oneside]{book} 

\usepackage[T1]{fontenc}
\usepackage[utf8]{inputenc}

\usepackage{ae}               
\usepackage[pdftex]{graphicx}
\usepackage{vmargin}          
\usepackage{fancyhdr}         
\usepackage{subfigure}
\usepackage[hyphens]{url}
\usepackage{xcolor}
\usepackage{footnote}

\usepackage[absolute,overlay]{textpos}
\usepackage{tikz}
\usepackage[english]{babel}
\usepackage{color}
\usepackage{longtable}
\usepackage[bottom]{footmisc}
\usepackage[section,subsection,subsubsection]{extraplaceins}
\usepackage{multirow}
\usepackage{multicol}	
\usepackage{array}
\newcolumntype{P}[1]{>{\raggedright\arraybackslash}p{#1}}
\usepackage{nomencl}
\usepackage{color}
\usepackage{setspace}
\usepackage{amsmath}
\usepackage{amssymb}

\makenomenclature

\makeglossary

\definecolor{darkred}{rgb}{0.5,0,0}
\definecolor{darkgreen}{rgb}{0,0.8,0}
\definecolor{darkblue}{rgb}{0,0,0.5}

\setmarginsrb
{2.5cm}   
{1.5cm}   
{2.5cm}   
{2cm}   
{20pt}    
{0.25in}  
{9pt}     
{0.3in}   

\usepackage{chngcntr}
\counterwithin{figure}{section} 
\counterwithin{table}{section}  

\newcommand{\forget}[1]{}  
\newcommand{\sandbox}[1]{}  
\newcommand{\longer}[1]{}  

\newcommand{\shorten}[1]{}  

\usepackage{authblk}

\author[1]{Tentative author list\footnote{SPEC Confidential. This document should be considered confidential unless labeled otherwise.}}

\usepackage{multirow}
\usepackage{array}
\usepackage{ragged2e}
\usepackage{hyperref}
\usepackage{natbib}

\newcommand\mytitle{Ready for Rain? A View from SPEC Research \\on the Future of Cloud Metrics}
\title{Cloud Metrics}

\newcommand\TRnumber{Technical Report: SPEC-RG-2016-01\\Version: 1.0}
\newcommand\WGname{SPEC RG Cloud Working Group}

\newcommand\TRdate{\today}
\newcommand\TRcentralURL{research.spec.org}
\newcommand\TRrightURL{www.spec.org}

\newcommand\Acknowledgements{The authors would like to thank Kai Sachs, Klaus Lange, and Manoj Karunakaran Nambiar}

\newcommand\numAuthors{7} 
\makeatletter
\newcommand\defcase[1]{\@namedef{mycase@\the\numexpr#1\relax}}
\newcommand\putAuthors[1]{\@nameuse{mycase@\the\numexpr#1\relax}}
\makeatother

\defcase{0}{
	\begin{textblock}{8}[0,0](\colsinglecentralX,\rowoneY)
		\centering
		\small{\textbf{No Author}}\\
	\end{textblock}	
}

\defcase{1}{
	\begin{textblock}{8}[0,0](\colsinglecentralX,\rowoneY)
		\centering
		\small{\textbf{\authorOneName}}\\
		\footnotesize
		\authorOneAffil
	\end{textblock}	
}
\defcase{2}{
	\begin{textblock}{\authorCellWidth}[0,0](\colDoubleLeftX,\rowoneY)
		\centering
		\small{\textbf{\authorOneName}}\\
		\footnotesize
		\authorOneAffil
	\end{textblock}	
	\begin{textblock}{\authorCellWidth}[0,0](\colDoubleRightX,\rowoneY)
		\centering
		\small{\textbf{\authorTwoName}}\\
		\footnotesize
		\authorTwoAffil
	\end{textblock}	
}
\defcase{3}{
	\begin{textblock}{\authorCellWidth}[0,0](\coloneX,\rowoneY)
		\centering
		\small{\textbf{\authorOneName}}\\
		\footnotesize
		\authorOneAffil
	\end{textblock}	
	\begin{textblock}{\authorCellWidth}[0,0](\coltwoX,\rowoneY)
		\centering
		\small{\textbf{\authorTwoName}}\\
		\footnotesize
		\authorTwoAffil
	\end{textblock}	
	\begin{textblock}{\authorCellWidth}[0,0](\colthreeX,\rowoneY)
		\centering
		\small{\textbf{\authorThreeName}}\\
		\footnotesize
		\authorThreeAffil
	\end{textblock}	
}
\defcase{4}{
	\begin{textblock}{\authorCellWidth}[0,0](\coloneX,\rowoneY)
		\centering
		\small{\textbf{\authorOneName}}\\
		\footnotesize
		\authorOneAffil
	\end{textblock}	
	\begin{textblock}{\authorCellWidth}[0,0](\coltwoX,\rowoneY)
		\centering
		\small{\textbf{\authorTwoName}}\\
		\footnotesize
		\authorTwoAffil
	\end{textblock}	
	\begin{textblock}{\authorCellWidth}[0,0](\colthreeX,\rowoneY)
		\centering
		\small{\textbf{\authorThreeName}}\\
		\footnotesize
		\authorThreeAffil
	\end{textblock}
	\begin{textblock}{8}[0,0](\colsinglecentralX,\rowtwoY)
		\centering
		\small{\textbf{\authorFourName}}\\
		\footnotesize
		\authorFourAffil
	\end{textblock}	
}
\defcase{5}{
	\begin{textblock}{\authorCellWidth}[0,0](\coloneX,\rowoneY)
		\centering
		\small{\textbf{\authorOneName}}\\
		\footnotesize
		\authorOneAffil
	\end{textblock}	
	\begin{textblock}{\authorCellWidth}[0,0](\coltwoX,\rowoneY)
		\centering
		\small{\textbf{\authorTwoName}}\\
		\footnotesize
		\authorTwoAffil
	\end{textblock}	
	\begin{textblock}{\authorCellWidth}[0,0](\colthreeX,\rowoneY)
		\centering
		\small{\textbf{\authorThreeName}}\\
		\footnotesize
		\authorThreeAffil
	\end{textblock}
	\begin{textblock}{\authorCellWidth}[0,0](\colDoubleLeftX,\rowtwoY)
		\centering
		\small{\textbf{\authorFourName}}\\
		\footnotesize
		\authorFourAffil
	\end{textblock}	
	\begin{textblock}{\authorCellWidth}[0,0](\colDoubleRightX,\rowtwoY)
		\centering
		\small{\textbf{\authorFiveName}}\\
		\footnotesize
		\authorFiveAffil
	\end{textblock}	
}
\defcase{6}{
	\begin{textblock}{\authorCellWidth}[0,0](\coloneX,\rowoneY)
		\centering
		\small{\textbf{\authorOneName}}\\
		\footnotesize
		\authorOneAffil
	\end{textblock}	
	\begin{textblock}{\authorCellWidth}[0,0](\coltwoX,\rowoneY)
		\centering
		\small{\textbf{\authorTwoName}}\\
		\footnotesize
		\authorTwoAffil
	\end{textblock}	
	\begin{textblock}{\authorCellWidth}[0,0](\colthreeX,\rowoneY)
		\centering
		\small{\textbf{\authorThreeName}}\\
		\footnotesize
		\authorThreeAffil
	\end{textblock}
	\begin{textblock}{\authorCellWidth}[0,0](\coloneX,\rowtwoY)
		\centering
		\small{\textbf{\authorFourName}}\\
		\footnotesize
		\authorFourAffil
	\end{textblock}	
	\begin{textblock}{\authorCellWidth}[0,0](\coltwoX,\rowtwoY)
		\centering
		\small{\textbf{\authorFiveName}}\\
		\footnotesize
		\authorFiveAffil
	\end{textblock}	
	\begin{textblock}{\authorCellWidth}[0,0](\colthreeX,\rowtwoY)
		\centering
		\small{\textbf{\authorSixName}}\\
		\footnotesize
		\authorSixAffil
	\end{textblock}
}
\defcase{7}{
	\begin{textblock}{\authorCellWidth}[0,0](\coloneX,\rowoneY)
		\centering
		\small{\textbf{\authorOneName}}\\
		\footnotesize
		\authorOneAffil
	\end{textblock}	
	\begin{textblock}{\authorCellWidth}[0,0](\coltwoX,\rowoneY)
		\centering
		\small{\textbf{\authorTwoName}}\\
		\footnotesize
		\authorTwoAffil
	\end{textblock}	
	\begin{textblock}{\authorCellWidth}[0,0](\colthreeX,\rowoneY)
		\centering
		\small{\textbf{\authorThreeName}}\\
		\footnotesize
		\authorThreeAffil
	\end{textblock}
	\begin{textblock}{\authorCellWidth}[0,0](\coloneX,\rowtwoY)
		\centering
		\small{\textbf{\authorFourName}}\\
		\footnotesize
		\authorFourAffil
	\end{textblock}	
	\begin{textblock}{\authorCellWidth}[0,0](\coltwoX,\rowtwoY)
		\centering
		\small{\textbf{\authorFiveName}}\\
		\footnotesize
		\authorFiveAffil
	\end{textblock}	
	\begin{textblock}{\authorCellWidth}[0,0](\colthreeX,\rowtwoY)
		\centering
		\small{\textbf{\authorSixName}}\\
		\footnotesize
		\authorSixAffil
	\end{textblock}
	\begin{textblock}{8}[0,0](\colsinglecentralX,\rowthreeY)
		\centering
		\small{\textbf{\authorSevenName}}\\
		\footnotesize
		\authorSevenAffil
	\end{textblock}
}
\defcase{8}{
	\begin{textblock}{\authorCellWidth}[0,0](\coloneX,\rowoneY)
		\centering
		\small{\textbf{\authorOneName}}\\
		\footnotesize
		\authorOneAffil
	\end{textblock}	
	\begin{textblock}{\authorCellWidth}[0,0](\coltwoX,\rowoneY)
		\centering
		\small{\textbf{\authorTwoName}}\\
		\footnotesize
		\authorTwoAffil
	\end{textblock}	
	\begin{textblock}{\authorCellWidth}[0,0](\colthreeX,\rowoneY)
		\centering
		\small{\textbf{\authorThreeName}}\\
		\footnotesize
		\authorThreeAffil
	\end{textblock}
	\begin{textblock}{\authorCellWidth}[0,0](\coloneX,\rowtwoY)
		\centering
		\small{\textbf{\authorFourName}}\\
		\footnotesize
		\authorFourAffil
	\end{textblock}	
	\begin{textblock}{\authorCellWidth}[0,0](\coltwoX,\rowtwoY)
		\centering
		\small{\textbf{\authorFiveName}}\\
		\footnotesize
		\authorFiveAffil
	\end{textblock}	
	\begin{textblock}{\authorCellWidth}[0,0](\colthreeX,\rowtwoY)
		\centering
		\small{\textbf{\authorSixName}}\\
		\footnotesize
		\authorSixAffil
	\end{textblock}
	\begin{textblock}{\authorCellWidth}[0,0](\colDoubleLeftX,\rowthreeY)
		\centering
		\small{\textbf{\authorSevenName}}\\
		\footnotesize
		\authorSevenAffil
	\end{textblock}
	\begin{textblock}{\authorCellWidth}[0,0](\colDoubleRightX,\rowthreeY)
		\centering
		\small{\textbf{\authorEightName}}\\
		\footnotesize
		\authorEightAffil
	\end{textblock}
}
\defcase{9}{
	\begin{textblock}{\authorCellWidth}[0,0](\coloneX,\rowoneY)
		\centering
		\small{\textbf{\authorOneName}}\\
		\footnotesize
		\authorOneAffil
	\end{textblock}	
	\begin{textblock}{\authorCellWidth}[0,0](\coltwoX,\rowoneY)
		\centering
		\small{\textbf{\authorTwoName}}\\
		\footnotesize
		\authorTwoAffil
	\end{textblock}	
	\begin{textblock}{\authorCellWidth}[0,0](\colthreeX,\rowoneY)
		\centering
		\small{\textbf{\authorThreeName}}\\
		\footnotesize
		\authorThreeAffil
	\end{textblock}
	\begin{textblock}{\authorCellWidth}[0,0](\coloneX,\rowtwoY)
		\centering
		\small{\textbf{\authorFourName}}\\
		\footnotesize
		\authorFourAffil
	\end{textblock}	
	\begin{textblock}{\authorCellWidth}[0,0](\coltwoX,\rowtwoY)
		\centering
		\small{\textbf{\authorFiveName}}\\
		\footnotesize
		\authorFiveAffil
	\end{textblock}	
	\begin{textblock}{\authorCellWidth}[0,0](\colthreeX,\rowtwoY)
		\centering
		\small{\textbf{\authorSixName}}\\
		\footnotesize
		\authorSixAffil
	\end{textblock}
	\begin{textblock}{\authorCellWidth}[0,0](\coloneX,\rowthreeY)
		\centering
		\small{\textbf{\authorSevenName}}\\
		\footnotesize
		\authorSevenAffil
	\end{textblock}	
	\begin{textblock}{\authorCellWidth}[0,0](\coltwoX,\rowthreeY)
		\centering
		\small{\textbf{\authorEightName}}\\
		\footnotesize
		\authorEightAffil
	\end{textblock}	
	\begin{textblock}{\authorCellWidth}[0,0](\colthreeX,\rowthreeY)
		\centering
		\small{\textbf{\authorNineName}}\\
		\footnotesize
		\authorNineAffil
	\end{textblock}
}
\defcase{10}{
	\begin{textblock}{\authorCellWidth}[0,0](\coloneX,\rowoneY)
		\centering
		\small{\textbf{\authorOneName}}\\
		\footnotesize
		\authorOneAffil
	\end{textblock}	
	\begin{textblock}{\authorCellWidth}[0,0](\coltwoX,\rowoneY)
		\centering
		\small{\textbf{\authorTwoName}}\\
		\footnotesize
		\authorTwoAffil
	\end{textblock}	
	\begin{textblock}{\authorCellWidth}[0,0](\colthreeX,\rowoneY)
		\centering
		\small{\textbf{\authorThreeName}}\\
		\footnotesize
		\authorThreeAffil
	\end{textblock}
	\begin{textblock}{\authorCellWidth}[0,0](\coloneX,\rowtwoY)
		\centering
		\small{\textbf{\authorFourName}}\\
		\footnotesize
		\authorFourAffil
	\end{textblock}	
	\begin{textblock}{\authorCellWidth}[0,0](\coltwoX,\rowtwoY)
		\centering
		\small{\textbf{\authorFiveName}}\\
		\footnotesize
		\authorFiveAffil
	\end{textblock}	
	\begin{textblock}{\authorCellWidth}[0,0](\colthreeX,\rowtwoY)
		\centering
		\small{\textbf{\authorSixName}}\\
		\footnotesize
		\authorSixAffil
	\end{textblock}
	\begin{textblock}{\authorCellWidth}[0,0](\coloneX,\rowthreeY)
		\centering
		\small{\textbf{\authorSevenName}}\\
		\footnotesize
		\authorSevenAffil
	\end{textblock}	
	\begin{textblock}{\authorCellWidth}[0,0](\coltwoX,\rowthreeY)
		\centering
		\small{\textbf{\authorEightName}}\\
		\footnotesize
		\authorEightAffil
	\end{textblock}	
	\begin{textblock}{\authorCellWidth}[0,0](\colthreeX,\rowthreeY)
		\centering
		\small{\textbf{\authorNineName}}\\
		\footnotesize
		\authorNineAffil
	\end{textblock}
	\begin{textblock}{8}[0,0](\colsinglecentralX,\rowfourY)
		\centering
		\small{\textbf{\authorTenName}}\\
		\footnotesize
		\authorTenAffil
	\end{textblock}
}

\newcommand\authorOneName{Nikolas Herbst}
\newcommand\authorOneAffil{
Chair for Software Engineering\\
	University of W\"urzburg\\
    W\"urzburg, Germany\\
	\emph{nikolas.herbst@uni-wuerzburg.de}
}

\newcommand\authorTwoName{Rouven Krebs}
\newcommand\authorTwoAffil{
	SAP AG\\
	Walldorf, Germany\\
	\emph{rouven.krebs@sap.com}
}
\newcommand\authorThreeName{Giorgos Oikonomou}
\newcommand\authorThreeAffil{
	Faculty of Engineering, Mathematics and Computer Science \\
	Delft University of Technology\\
	Delft, Netherlands\\
	\emph{g.oikonomou@student.tudelft.nl}
}
\newcommand\authorSixName{Alexandru Iosup}
\newcommand\authorSixAffil{
	Faculty of Engineering, Mathematics and Computer Science \\
	Delft University of Technology\\
	Delft, Netherlands\\
	\emph{A.Iosup@tudelft.nl}
}

\newcommand\authorSevenName{Samuel Kounev}
\newcommand\authorSevenAffil{
Chair for Software Engineering\\
	University of W\"urzburg\\
    W\"urzburg\\
	\emph{samuel.kounev@uni-wuerzburg.de}
}

\newcommand\authorFourName{George Kousiouris}
\newcommand\authorFourAffil{
    School of Electrical and Computer Engineering\\
	National Technical University of Athens\\
    Athens, Greece\\
	\emph{gkousiou@mail.ntua.gr}
}

\newcommand\authorFiveName{Athanasia Evangelinou}
\newcommand\authorFiveAffil{
	School of Electrical and Computer Engineering\\
	National Technical University of Athens\\
	Athens, Greece\\
	\emph{aevang@mail.ntua.gr}
}

\begin{document}
 
\selectlanguage{english} 
\frontmatter

\thispagestyle{empty}
\newcommand{\changefont}[3]{\fontfamily{#1} \fontseries{#2} \fontshape{#3} \selectfont}
\newcommand{\diameter}{20}
\newcommand{\xone}{-25}
\newcommand{\xtwo}{165}
\newcommand{\yone}{20}
\newcommand{\ytwo}{-253}

\newcommand{\rowoneY}{5.5}		
\newcommand{\rowtwoY}{7.0}
\newcommand{\rowthreeY}{8.5}
\newcommand{\rowfourY}{10.1}

\newcommand{\coloneX}{2.5}
\newcommand{\coltwoX}{7.45}
\newcommand{\colthreeX}{12.4}

\newcommand{\colDoubleLeftX}{5}
\newcommand{\colDoubleRightX}{10}

\newcommand{\colsinglecentralX}{5.9}

\newcommand{\authorCellWidth}{4.9}

\begin{titlepage}
\begin{tikzpicture}[overlay]
\draw[color=gray]  
 (\xone mm, \yone mm) -- (\xtwo mm, \yone mm) arc (90:0:\diameter pt) 
  -- (\xtwo mm + \diameter pt , \ytwo mm) -- (\xone mm + \diameter pt , \ytwo mm) 
 arc (270:180:\diameter pt) -- (\xone mm, \yone mm);
\end{tikzpicture}

\changefont{phv}{m}{n}	

\begin{textblock}{14}[0,0](3,2.3)
	\centering
	\large{\TRnumber}\\
	\vspace*{1cm}
	\huge{\mytitle}\\
	\vspace*{0.5cm}
	\Large{\WGname}
\end{textblock}
\begin{textblock}{15.5}[0,0](2,5.2)
	\begin{tikzpicture}
		\fill[red!80!brown] (0,0cm) rectangle (19.5cm,0.1cm);
	\end{tikzpicture}
\end{textblock}

\begin{center}
	\putAuthors{\numAuthors}
\end{center}

\begin{textblock}{14}[0,0](3,13)
	\hfill
	\includegraphics[width=3cm]{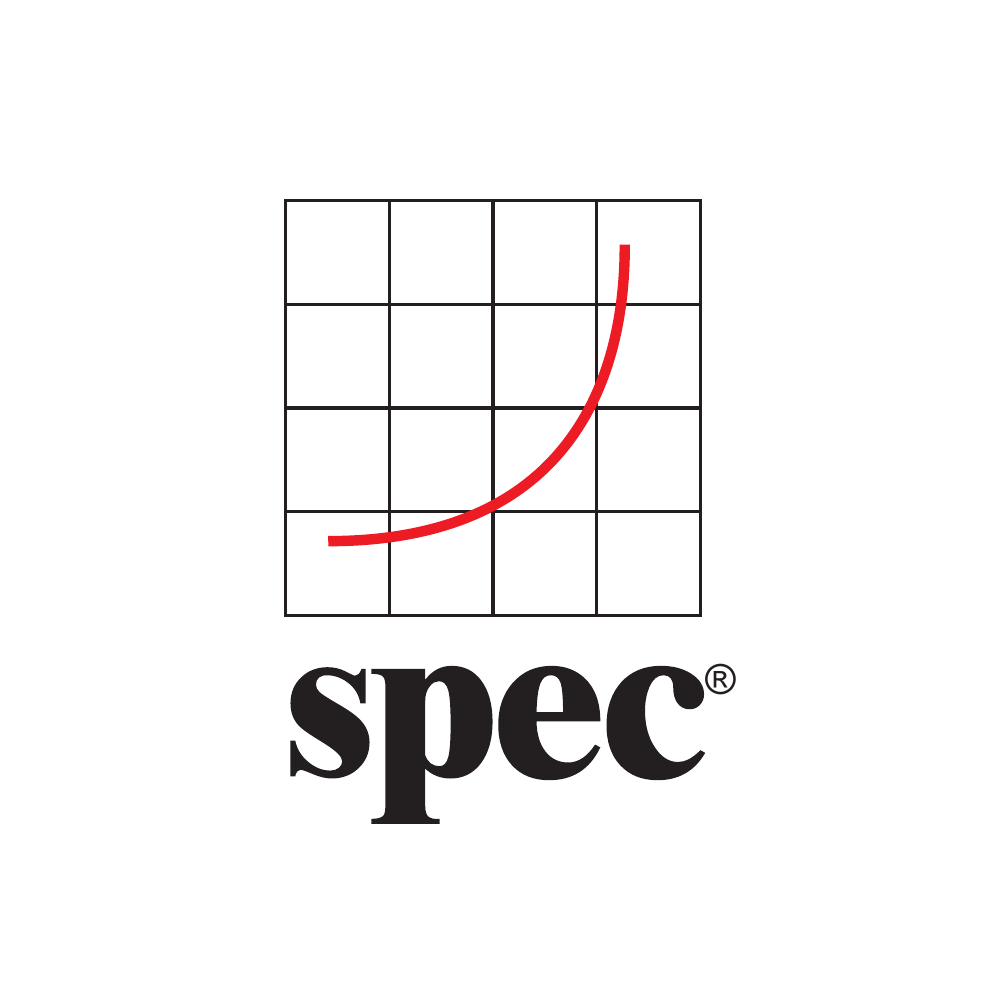} \hfill
	\includegraphics[width=1.9cm]{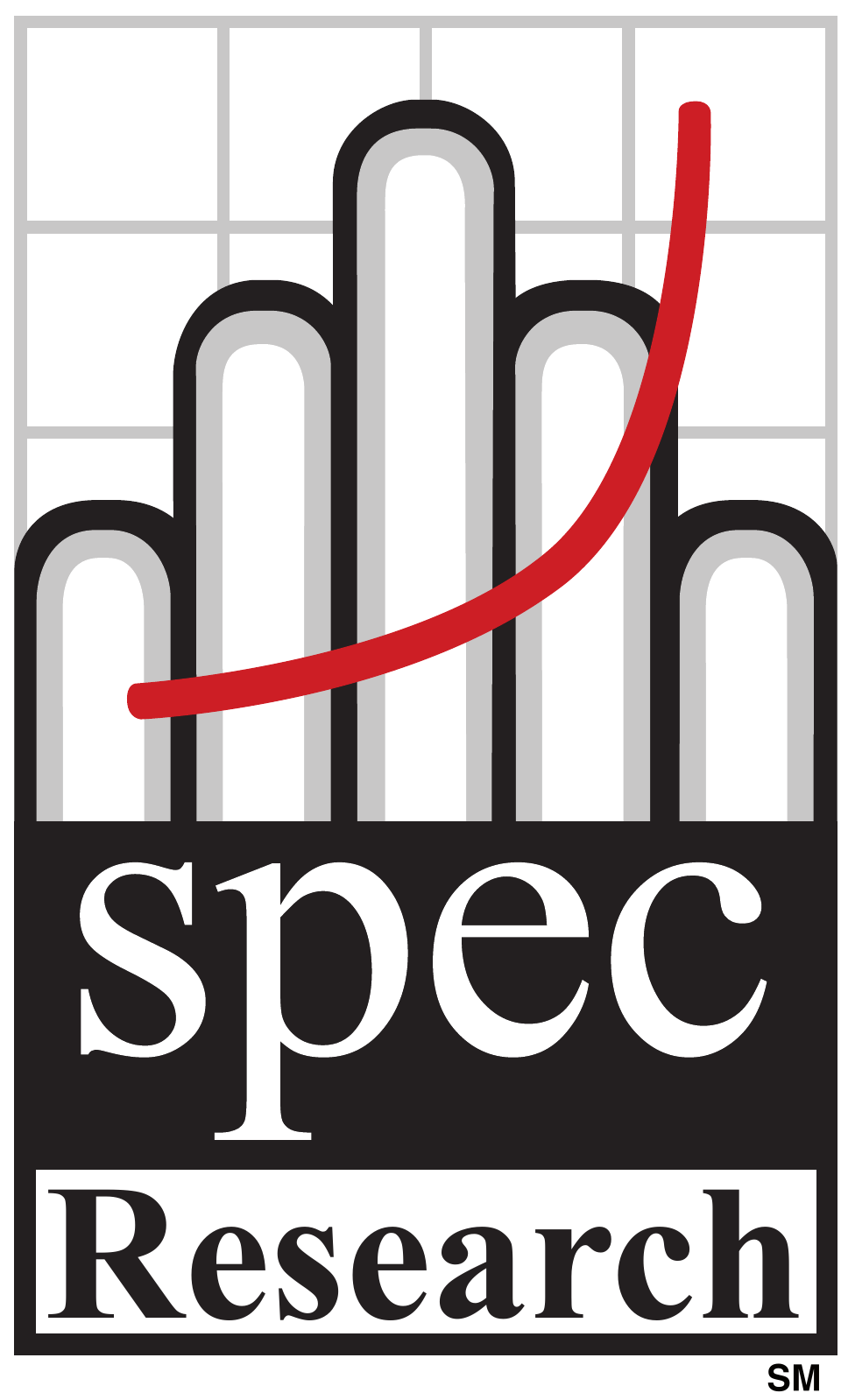} \hspace{1.1cm}\hfill
	\includegraphics[width=1.9cm]{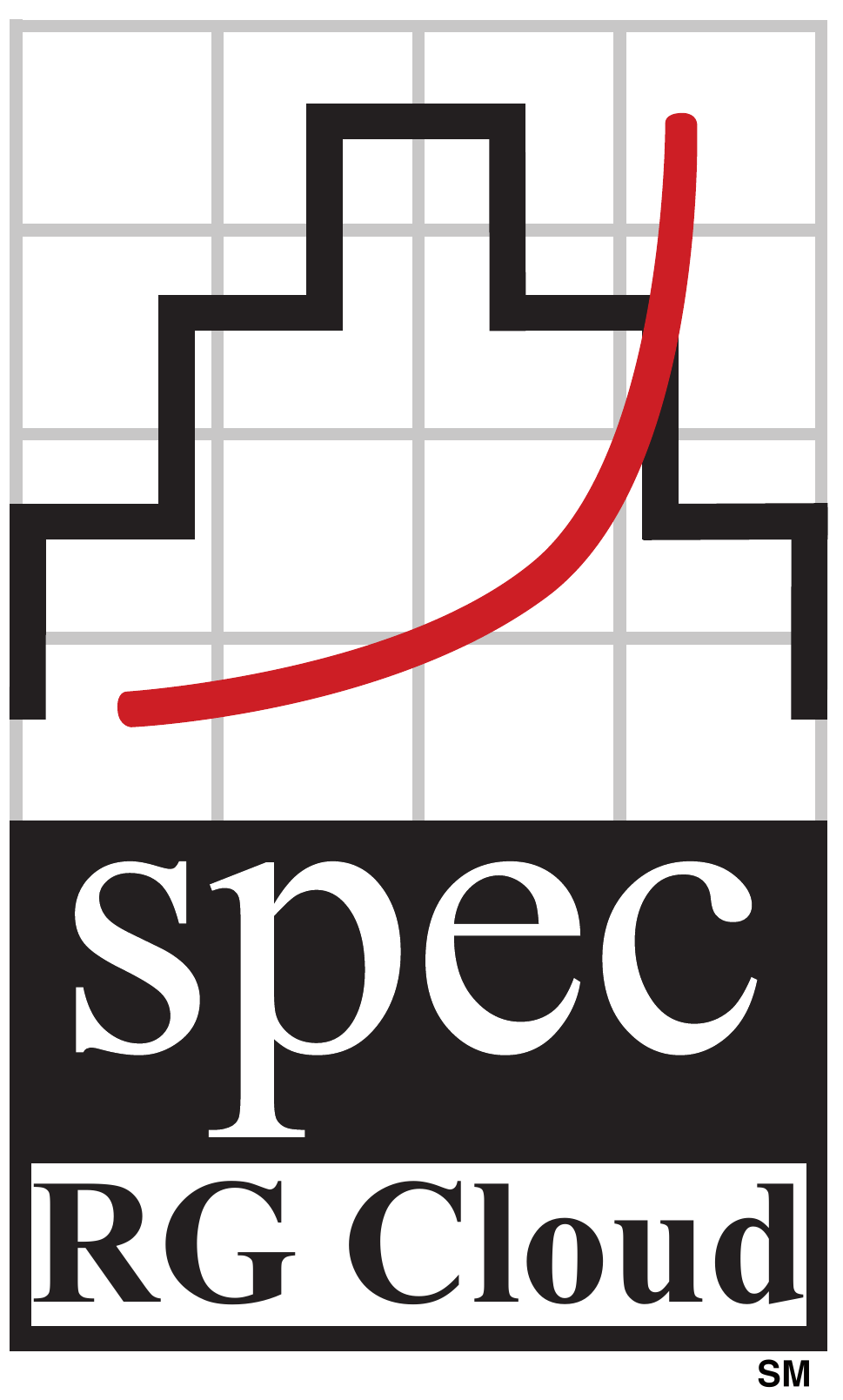} \hspace{1.1cm}\hfill
	\hfill
\end{textblock}

\begin{textblock}{14}[0,0](3,15)
	\noindent\textbf{Acknowledgements}\hfill\vspace{0.5em}\hrule
	\vspace{0.5em}\noindent\footnotesize
	\Acknowledgements
\end{textblock}

\begin{textblock}{14}[0,0](3,16.75)
	\centering
	\large{\textbf{\TRdate}}
	\hfill
	\large{\textbf{\TRcentralURL}}
	\hfill
	\large{\textbf{\TRrightURL}}
\end{textblock}

\end{titlepage}

\newpage 
\thispagestyle{empty}
\mbox{}

\newpage
\pagenumbering{roman}
\setcounter{tocdepth}{4}
\begin{spacing}{1.3}
\tableofcontents
\end{spacing}

\newpage
\thispagestyle{plain}
\section*{Executive Summary}

In the past decade, cloud computing has emerged from a pursuit for a service-driven information and communication technology (ICT), into a significant fraction of the ICT market.
Responding to the growth of the market, many alternative cloud services and their underlying systems are currently vying for the attention of cloud users and providers. 
Thus, benchmarking them is needed, to enable cloud users to make an informed choice,
and to enable system DevOps to tune, design, and evaluate their systems.
This requires focusing on old and new system properties, possibly leading to the re-design of classic benchmarking metrics, such as expressing performance as throughput and latency (response time), and the design of new, cloud-specific metrics.

Addressing this requirement, in this work we focus on four system properties: 

(i) \emph{elasticity} of the cloud service, to accommodate large variations in the amount of service requested, 

(ii) \emph{performance isolation} between the tenants of shared cloud systems, 

(iii) \emph{availability} of cloud services and systems, and the 

(iv) \emph{operational risk} of running a production system in a cloud environment.  

\noindent Focusing on key metrics, for each of these properties 
we review the state-of-the-art, then select or propose new metrics together with measurement approaches. 
We see the presented metrics as a foundation towards upcoming, industry-standard, cloud benchmarks.

\vspace{0.5cm}

\textbf{Keywords\footnote{The used keywords are defined as part of The 2012 ACM Computing Classification System \cite{acm:classification}.}:} \\

\noindent Cloud Computing; Metrics; Measurement; Benchmarking; Elasticity; Isolation; Performance; Service Level Objective; Availability; Operational Risk\\

\noindent CCS -  General and reference - Metrics;\\
CCS -  General and reference - Surveys and overviews;\\
CCS -  Computer systems organization -  Architectures - \newline 
Distributed architectures - Cloud computing;\\
CCS -  Software and its engineering - Extra-functional properties. \\

\textbf{Trademark}\\

\noindent SPEC, the SPEC logo and the name SPEC CPU2006 are trademarks of the Standard Performance Evaluation Corporation (SPEC). SPEC Research and SPEC RG Cloud are servicemarks of SPEC.  Additional product and service names mentioned herein may be the trademarks of their respective owners.
Copyright Notice
Copyright © 1988-2016 Standard Performance Evaluation Corporation (SPEC). All rights reserved.

\mainmatter
\section{Introduction}

Cloud computing is a paradigm under which ICT services are offered ``as a service'', that is, on-demand, and with payment expected to match what is actually used. Over the last decade, cloud computing has become increasingly important for the information and communication technology (ICT) industry. 
Cloud applications already represent over 10\% of the entire ICT market in Europe~\cite{ec/CloudUptake14}, and likely a similarly significant fraction of the ICT market in North America, Middle East, and Asia.
By 2017, over three-quarters of the global business and personal data may reside in cloud data-centers, according to a recent IDC report~\cite{IDC15}. 
This promising growth trend makes clouds an interesting new target for benchmarking, with the goal of comparing, tuning, and improving the increasingly large set of cloud-based systems and applications, and the cloud fabric itself.
However, traditional benchmarking approaches may not be able to address the new cloud computing settings.
In classical benchmarking, common system performance metrics are measured on well-defined, often well-behaved systems-under-test (SUTs).
In contrast, cloud systems can be built out of a rich, yet volatile combination of infrastructure, platforms, and entire software stacks, which in turn can be built out of cloud systems and offered as cloud services.
For example, Netflix currently streams video as a cloud service to millions of people world-wide, occupying a significant fraction of the download capacity of Internet home-users, and, simultaneously, uses the infrastructure and the content-distribution platform provided as cloud services by Amazon AWS. 
Key to benchmarking the rich tapestry that characterizes many cloud services and their underlying systems is the re-definition of traditional benchmarking metrics for cloud settings, and the definition of new metrics that are unique to cloud computing. This is the focus of our work, and the main contribution of this report.

Academic studies, concerned public reports, and even company white papers indicate that 
a variety of new operational and user-driven phenomena take place in cloud settings.
We consider in this work four such phenomena.
First, cloud systems have been 
asked to deliver an illusion of infinite capacity and capability, raising interesting questions of how to provide results under wildly varying workloads, and forcing cloud systems to appear perfectly {\it elastic}. 
Second, cloud services and systems have been shown to exhibit high performance variability~\cite{Iosup2010}, against which modern cloud users have requested protection ({\it performance isolation}).
Third, increasingly more demanding users expect today that the {\it availability} of cloud services is nearly perfect, and even few unavailability events can cause significant reputation and pecuniary damage to a cloud provider. 
Fourth, as the risks of not meeting implicit user-expectations and explicit service contracts (service level agreements, SLAs) are increasing with the scale of cloud operations, cloud providers have become increasingly more interested to reduce their {\it operational risk}.

With the market growing and maturing, many cloud services are now competing for the attention of existing and new cloud users. 
Thus, quantifying the capabilities of the system features that respond to various cloud phenomena, and in particular benchmarking the non-functional properties of cloud systems (including performance), is increasingly important.  
We ask in this work three important research questions: 
{\it Do traditional metrics already support the cloud features created to match the four phenomena we consider in this work?}
As we report in this work, the survey of the state-of-the-art indicates that the answer to this question is ``No.'', we further raise the follow-up research question: {\it Which new metrics are needed, to support the cloud features created to match the four phenomena we consider in this work?} To this question we cannot offer a definitive answer, but our results include various new metrics and adaptations of existing metrics that may lead to new industrial-grade benchmarks.

Addressing the two research questions, the goal of this report is to lay a foundation for making various cloud offerings and technologies comparable to each other, and provide a common understanding among all cloud stakeholders. 
Although elasticity, isolation, availability, and operational risk are already perceived as important aspects in the academia and by the industry, they have never before been thoroughly defined and surveyed. As we show in this work, their meaning can be different for different stakeholders, and in some cases existing definitions are inconsistent or even contrary to each other. 
Toward reaching our goal, our main contribution is four-fold. Each contribution is focusing on the foundations of benchmarking one feature of cloud systems. In turn, the features are:
\begin{enumerate}
	\item {\it Elasticity}, addressed in Section~\ref{sec:Elasticity}.
	Elasticity offers the opportunity to automatically adapt the resource supply to a changing demand. 
The quality of elastic adaptation is only indirectly captured by traditional performance metrics, such as response time and utilization, and requires new approaches.
We present in this work a set of metrics and methods for combining them to capture the accuracy and timing aspects of elastic platforms.

	\item {\it Performance isolation}, addressed in Section~\ref{sec:PerfIsolation}.
	The underlying cloud infrastructure has the important task to isolate different customers sharing the same hardware from each other with regards to the performance they observe.
	We present in this work metrics that capture the influence of disruptive workloads, 
	the maximum disruptive load that a system can handle,
	and the degree of breaches of the performance isolation agreement.
	
	\item {\it Availability}, addressed in Section~\ref{sec:Availability}.
To quantify the availability of their business critical cloud applications and compare them for different contexts, in this work we analyze the availability definitions used by various cloud providers. We then define a simple metric of SLA adherence that enables direct comparisons between providers with otherwise different definitions of availability.

	\item {\it Operational risk-related}, addressed in Section~\ref{sec:Risk}.
	And on a more general level than the other features, we also focus in this work on estimating different types of operational risks that are connected with running software in the cloud. We define here various relevant metrics, and a measurement methodology that addresses them. 
	
\end{enumerate}

\section{Elasticity}
\label{sec:Elasticity}
\subsection{Goal and Relevance}

Elasticity has originally been defined in physics as a material property capturing the capability of 
returning to its original state after a deformation.
In economics, elasticity captures the effect of change in one variable to another dependent variable.
In both cases, elasticity is an intuitive concept and can be precisely described using mathematical formulas.
  
The concept of elasticity has been transferred to the context of cloud computing and 
is commonly considered as one of the central attributes of the cloud paradigm as in~\cite{gartnerReport}.  
For marketing purposes, the term elasticity is heavily used in cloud providers' advertisements 
and even in the naming of specific products or services.
Even though tremendous efforts are invested to enable cloud systems to behave in an elastic manner, 
no common and precise understanding of this term in the context of cloud computing has been established so far, 
and no ways have been proposed to quantify and compare elastic behavior.

\subsection{Prerequisites}
	
The scalability of a system including all hardware, virtualization, and software layers within 
its boundaries is a prerequisite in order to be able to speak of elasticity. 
Scalability is the ability of a system to sustain increasing workloads with adequate performance provided that hardware resources are added. 
Scalability in the context of distributed systems has been defined in \cite{Jogalekar00scalability}, as well as more recently in
\cite{Duboc2009PhDAFrameworkForTheCharacterizationAndAnalysisOfSoftwareSystemsScalability,
DubocRosenblumWicks2007AFrameworkForCharacterizationAndAnalysisOfSoftwareSystemScalability}, 
where also a measurement methodology is proposed. 

Given that elasticity is related to the ability of a system to adapt to changes in workloads and resource demands,
the existence of at least one specific adaptation process is assumed. 
The latter is normally automated, however, in a broader sense, it could also contain manual steps.
Without a defined adaptation process, a scalable system cannot behave in an elastic manner, 
as scalability on its own does not include temporal aspects.
                
When evaluating elasticity, the following points need to be checked beforehand:
\begin{itemize}
\item \emph{Autonomic Scaling:} \\What adaptation process is used for autonomic scaling?
\item \emph{Elasticity Dimensions:} \\What is the set of resource types scaled as part of the adaptation process?
\item \emph{Resource Scaling Units:} \\For each resource type, in what unit is the amount of allocated resources varied?
\item \emph{Scalability Bounds:} \\For each resource type, what is the upper bound on the amount of resources that can be allocated?
\end{itemize} 

\subsection{Definition}

\begin{description}
\item[Elasticity] is the degree to which a system is able to adapt to workload changes by provisioning and de-provisioning resources in an autonomic manner, 
such that at each point in time the available resources \emph{match} the current demand as closely as possible.
\end{description}

\paragraph*{Dimensions and Core Aspects}                

Any given adaptation process is defined in the context of at least one or possibly 
multiple types of resources that can be scaled up or down as part of the adaptation. 
Each resource type can be seen as a separate dimension of the adaptation process with its own elasticity properties.
If a resource type is a container of other resources types, like in the case of a virtual machine 
having assigned CPU cores and RAM, elasticity can be considered at multiple levels. 
Normally, resources of a given resource type can only be provisioned in discrete units like CPU cores, virtual machines (VMs), or physical nodes. 
For each dimension of the adaptation process with respect to a specific resource type, 
elasticity captures the following core aspects of the adaptation:
\begin{description}
\item [Timing] The timing aspect is captured by the percentages a system is in
an under-provisioned, over-provisioned or perfect state and by the amount of
superfluous adaptations.
\item [Accuracy] The accuracy of scaling is defined as the average absolute 
deviation of the current amount of allocated resources from the actual resource demand.
\end{description}

As discussed above, elasticity is always considered with respect to one or more resource types. 
Thus, a direct comparison between two systems in terms of elasticity 
is only possible if the same resource types (measured in identical units) are scaled. 

To evaluate the actual observable elasticity in a given scenario, 
as a first step, one must define the criterion based on which the amount of provisioned resources is considered to \emph{match} 
the actual current demand needed to satisfy the system's given performance requirements.
Based on such a matching criterion, specific metrics that quantify the above mentioned core aspects, as discussed in more detail in Section \ref{sec:metrics}, 
can be defined to quantify the practically achieved elasticity in comparison to the hypothetical \emph{optimal elasticity}. 
The latter corresponds to the hypothetical case where the system is scalable with respect to all considered elasticity dimensions 
without any upper bounds on the amount of resources that can be provisioned and where resources are provisioned and de-provisioned immediately as they are needed 
exactly matching the actual demand at any point in time. 
\emph{Optimal elasticity}, as defined here, would only be limited by the resource scaling units. 

\paragraph*{Differentiation}

This paragraph discusses the conceptual differences between elasticity and the related terms scalability and efficiency. 

\begin{description}
\item [Scalability] is a prerequisite for elasticity, but it does not consider temporal aspects of how fast, 
how often, and at what granularity scaling actions can be performed. 
Scalability is the ability of the system to sustain increasing workloads by making use of additional resources,
and therefore, in contrast to elasticity, it is not directly related to how well the actual resource demands are matched 
by the provisioned resources at any point in time. 
\item [Efficiency] expresses the amount of resources consumed for processing a given amount of work. 
In contrast to elasticity, efficiency is not limited to resource types that are scaled as part of the system's adaptation mechanisms. 
Normally, better elasticity results in higher efficiency. 
The other way round, this implication is not given, as efficiency can be influenced by other factors
independent of the system's elasticity mechanisms (e.g., different implementations of the same operation).  
\end{description}

\subsection{Derivation of the Matching Function}
\label{sec:DerivationMatching}

To capture the criterion based on which the amount of provisioned resources is considered to match the actual current demand, 
we define a matching function $m(w)=r$ as a system specific function that returns the minimal amount of resources $r$ for a given resource type 
needed to satisfy the system's performance requirements at a specified workload intensity. 
The workload intensity $w$ can be specified either as the number of workload 
units (e.g., user requests) present at the system at the same time (concurrency level), 
or as the number of workload units that arrive per unit of time (arrival rate). 
A matching function is needed for both directions of scaling (up/down), as it cannot be assumed that 
the optimal resource allocation level when transitioning from an under-provisioned state (upwards) 
are the same as when transitioning from an over-provisioned state (downwards).

\begin{figure}[htbp]
	\begin{center}
		\includegraphics[trim = 10mm 30mm 10mm 65mm,clip, width=0.8\columnwidth]{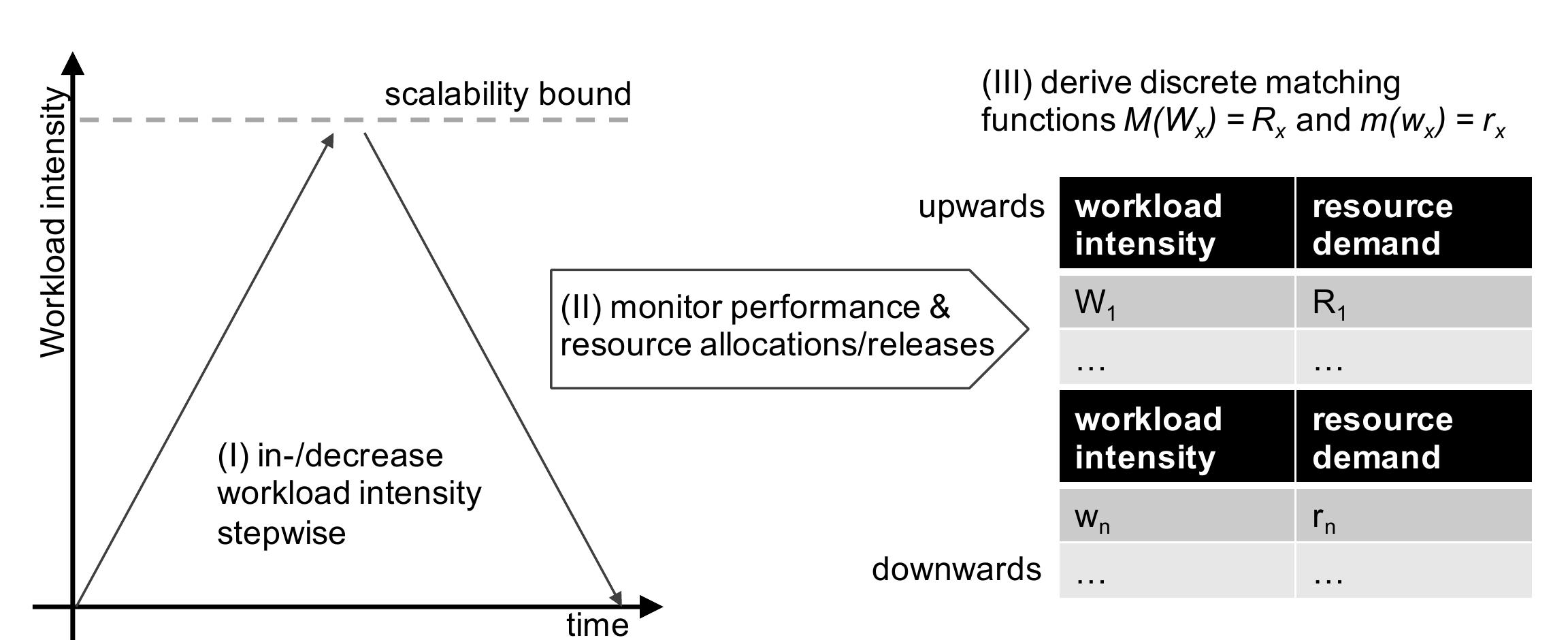}
		\caption{Illustration of a measurement-based derivation of matching functions}
		\label{fig:matchingfunction}
	\end{center}
\end{figure}

The matching functions can be derived based on measurements, as illustrated in Figure \ref{fig:matchingfunction},
by increasing the workload intensity $w$ stepwise, while measuring the resource consumption $r$,
and tracking resource allocation action changes. 
The process is then repeated for decreasing $w$.
After each change in the workload intensity, the system should be given enough time to adapt its resource allocations 
reaching a stable state for the respective workload intensity. 
As a result of this step, a system specific table is derived 
that maps workload intensity levels to resource demands, and the other way round, for both scaling directions within the scaling bounds.

\subsection{Related Elasticity Metrics}

In this section, we group existing metrics and benchmark approaches for
elasticity according to their perspective and discuss shortcomings.\\

\paragraph*{Elasticity Metrics:} 

Several metrics for elasticity have been
proposed so far:

\noindent (i) The ``scaling latency'' metrics in \cite{Li2010,Li2012} or the
``provisioning interval'' in \cite{Chandler2012} capture the time to bring up or drop a resource. This
duration is a technical property of elastic environments independent of
demand changes and the elasticity mechanism itself that decides when to trigger
a reconfiguration. Thus, these metrics are insufficient to fully characterize the
elasticity of a platform.\newline
(ii) The ``elastic speedup'' metric proposed by the SPEC OSG in
\cite{Chandler2012} relates the processing capability of a system at different
scaling levels. This metric does not capture any dynamic aspect of
elasticity and is regarded as scalability metric. \newline
(iii) The ``reaction time'' metric in \cite{KuHeKiRe2011-ResourceElasticity} can
only be computed if a unique mapping between resource demand changes and supply
changes exists. This assumption does not hold especially for proactive or
unstable elasticity mechanisms.\newline 
(iv) The approaches in \cite{Binnig2009, Cooper2010, Almeida13, Dory2011}
describe elasticity indirectly by analysing response times for significant
changes or SLO compliance. In theory, perfect elasticity results in constant
response times for varying arrival rates. In practice, detailed reasoning about
the quality of platform adaptations based on response times is hampered due to
 black-box abstractions, e.g., that the amount of surplus resources remains
hidden and individual requests have inherently different response times.\newline 
(v) Cost-based metrics are proposed in \cite{Islam2012, Folkerts2012,
Suleiman2012, Weinman2011, Tinnefeld2014} to quantify the impact of 
elasticity either by comparing the resulting costs to the costs for a peak-load
static assignment of resources or the costs of a hypothetical perfect elastic
platform. In both cases, the resulting metrics strongly depend on the
underlying cost model as well as on the assumed penalty for under-provisioning
and thus do not allow fair cross-platform comparison.\newline
(vi) The integral-based ``agility'' metric proposed by the SPEC OSG in
\cite{Chandler2012} compares the demand and supply over time normalized by
the average demand. They state the metric becomes invalid in cases where
SLOs are not met. This metric resembles the previously proposed ``precision''
metric in \cite{HeKoRe2013-ICAC-Elasticity} and is included in a refined version
normalized by time in this document (see Section \ref{ch:Metrics:sec:Accuracy}) to
capture the accuracy aspect of elastic adaptations also in situations when SLOs are not met.

\paragraph*{Benchmarks: } Existing benchmark approaches for elasticity as in
\cite{Folkerts2012, Suleiman2012, Weinman2011, Shawky12, Islam2012, Dory2011, Almeida13,
Tinnefeld2014, Cooper2010} account neither for differences in
efficiency of the underlying physical resources nor for possibly non-linear
scalability of the platform. As a consequence, the quantification of resource
elasticity is not realized in isolation of these related platform attributes as
previously highlighted in \cite{HeKoRe2013-ICAC-Elasticity}.
In contrast, our proposed approach uses the insights gained in an automated scalability
and performance analysis to adapt the load profile in a platform
specific way. 
In addition, existing benchmarks employ load profiles, which rarely cover the
realistic variability of the load intensity over time. In several cases,
the aspect of scaling downwards is omitted as in \cite{Dory2011, Shawky12}. In
\cite{Islam2012}, sinus like load profiles with plateaus are employed.
Real-world load profiles exhibit a mixture of seasonal patterns, trends, bursts
and noise. We account for the generic benchmark requirement
``representativeness'' as mentioned in~\cite{Huppler2009} by employing the load profile modeling
formalism DLIM presented in \cite{KiHeKo2014-LT-DLIM, KistowskiMA2014}.

\subsection{Proposed Elasticity Metrics}
\label{sec:metrics}

The \emph{demand} of a certain load intensity is understood as the minimal
amount of resources required for fulfilling a given performance related service
level objective (SLO).
The metrics are designed to characterize two core aspects of
elasticity: \emph{Accuracy} and \emph{timing}\footnote{In
\cite{HeKoRe2013-ICAC-Elasticity}, these aspects are referred to as
\emph{precision} and \emph{speed}.}. For all metrics, the optimal value is zero
and defines the perfect elastic system.
For a valid comparison of elasticity based on the proposed set
of metrics, the platforms (i) require the existence of an autonomic adaption
process, (ii) the scaling of the same resource type, e.g., CPU cores or virtual
machines (VMs) and (iii) within the same ranges, e.g., 1 to 20 resource units.

The metrics evaluate the resulting elastic behavior and thus are not designed
for distinct descriptions of the underlying hardware, the virtualization technology,
the used cloud management software or the used elasticity strategy and its
configuration. As a consequence, the metric and the measurement methodology are
applicable in situations where not all influencing factors are known.
All metrics require two discrete curves as input: The demand curve, which
defines how the resource demand varies during the measurement period, and the
supply curve, which defines how the amount of actually used resource varies.

The following Section~\ref{ch:Metrics:sec:Accuracy} describes the metrics for
the \emph{accuracy} aspect whereas Section \ref{ch:Metrics:sec:Timing} presents
a set of metrics for the quantification of the \emph{timing} aspect.
In Section~\ref{ch:Metrics:Compare}, we outline an approach for the aggregation of the proposed metrics enabling to compute a consistent ranking between
multiple elastic cloud environments and configurations.
\subsubsection{Accuracy}
\label{ch:Metrics:sec:Accuracy}

\begin{figure}[htb]
\begin{center}
  \includegraphics[width=0.8\columnwidth, trim=0cm 0cm 3cm
  6cm, clip=true]{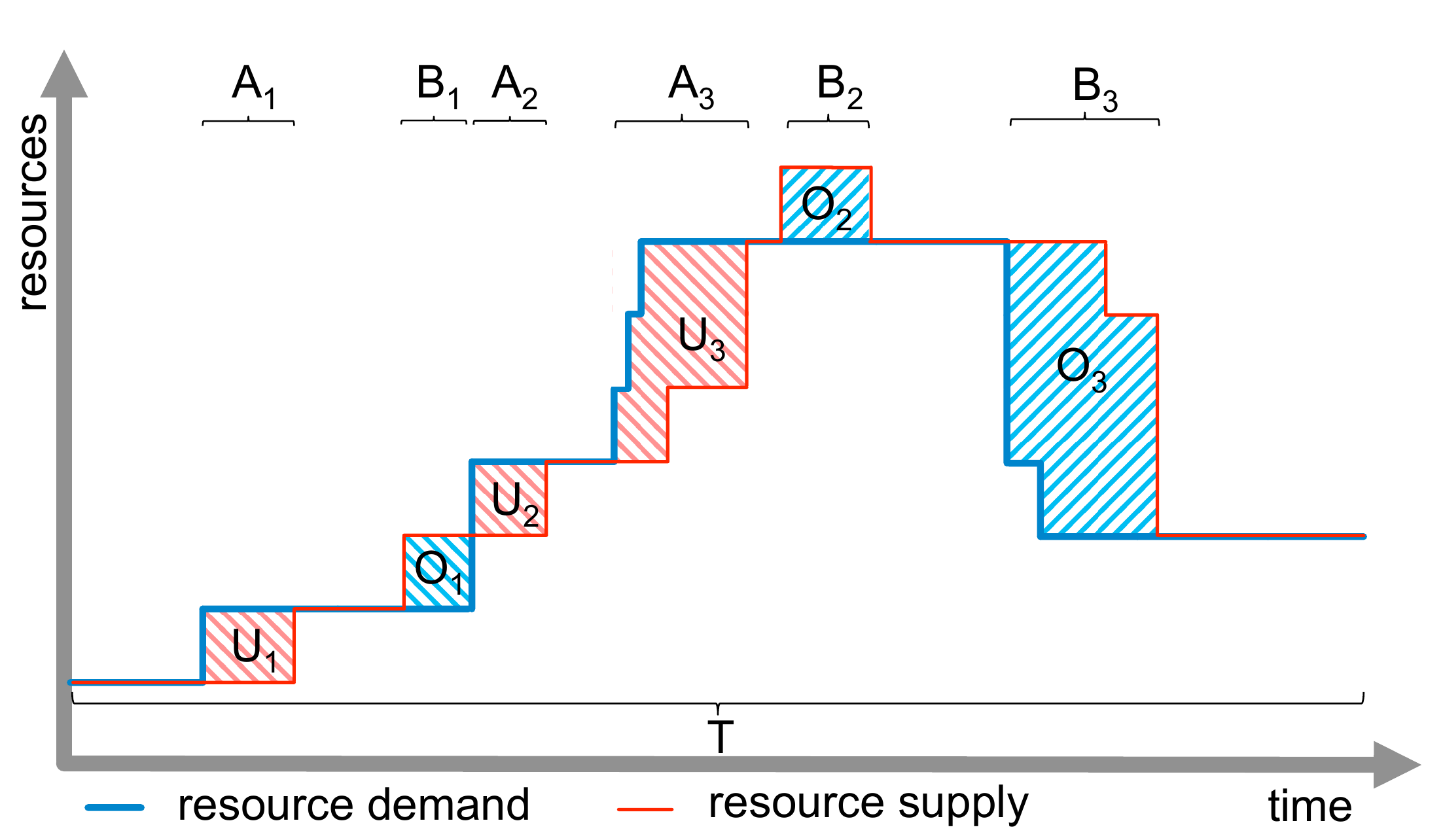} \caption[Measuring accuracy:
  red/blue areas indicate under-/over-provisioning]{Illustration for the
  definition of accuracy and provisioning timeshare metrics.}
  \label{fig:metrics}
\end{center}
\end{figure}

As visualized in Fig. \ref{fig:metrics}, the under-provisioning accuracy metric
$accuracy_U$ formally defined in the previous publication
\cite{HeKoRe2013-ICAC-Elasticity}, is calculated as the sum of areas $\sum{U}$
where the resource demand exceeds the supply normalized by the duration of the measurement period $T$. Accordingly, the over-provisioning accuracy metric $accuracy_O$ bases on the sum of areas $(\sum{O}$) where the resource supply exceeds the demand.

\begin{flalign}
  \text{Under-provisioning: } accuracy_U~[resource~units]
  &=  \frac{\sum{U}}{T}\\
  \text{Over-provisioning: } accuracy_O~[resource~units]
  &=  \frac{\sum{O}}{T}
\end{flalign}

\vspace{-4mm}
\vspace{-2mm}
Thus, $accuracy_U$ and $accuracy_O$ are the average of the absolute deviations between the current amounts of allocated resources and their respective actual resource demands. 
Since under-provisioning results in violating SLOs, a customer might want to use platform that do not
tend to under-provision at all. Thus, the challenge for providers is to ensure that enough resources are provided at
any point in time, but at the same time beat the competitors by
over-provisioning not too much. Considering this, separate accuracy measures for
over-provisioning and under-provisioning help providers to communicate their
elasticity capabilities and customers to select a cloud provider according
to their needs. 

\subsubsection{Timing}
\label{ch:Metrics:sec:Timing}
We highlight the \emph{timing} aspect of elasticity from the viewpoints of the
pure \emph{provisioning timeshare} and the \emph{jitter} accounting for superfluous or skipped adaptations.

\paragraph{Provisioning Timeshare}
The two accuracy metrics allow no reasoning whether the average amount of
under-/over-provisioned resources results from a few big deviations between
demand and supply or if it is rather caused by a constant small deviation. To
address this, the following two metrics are designed to give more insights about
the ratio of time in which under- or over-provisioning occurs.

As visualized in Fig. \ref{fig:metrics}, the following metrics $timeshare_U$ and
$timeshare_O$ are computed by summing up the total amount of time spend in an under- $(\sum{A})$ or over-provisioned
$(\sum{B})$ state normalized by the duration of the measurement period. Thus, they measure the overall timeshare spent in
under- or over-provisioned states:
\begin{flalign}
  \text{Under-provisioning: } timeshare_U~[\%]&= 
  \frac{\sum{A}}{T}\\
  \text{Over-provisioning: } timeshare_O~[\%]&= 
  \frac{\sum{B}}{T}
\end{flalign}

\begin{figure}[htb]
    \centering \subfigure[Platform A]{
   \includegraphics[width=0.8\columnwidth]{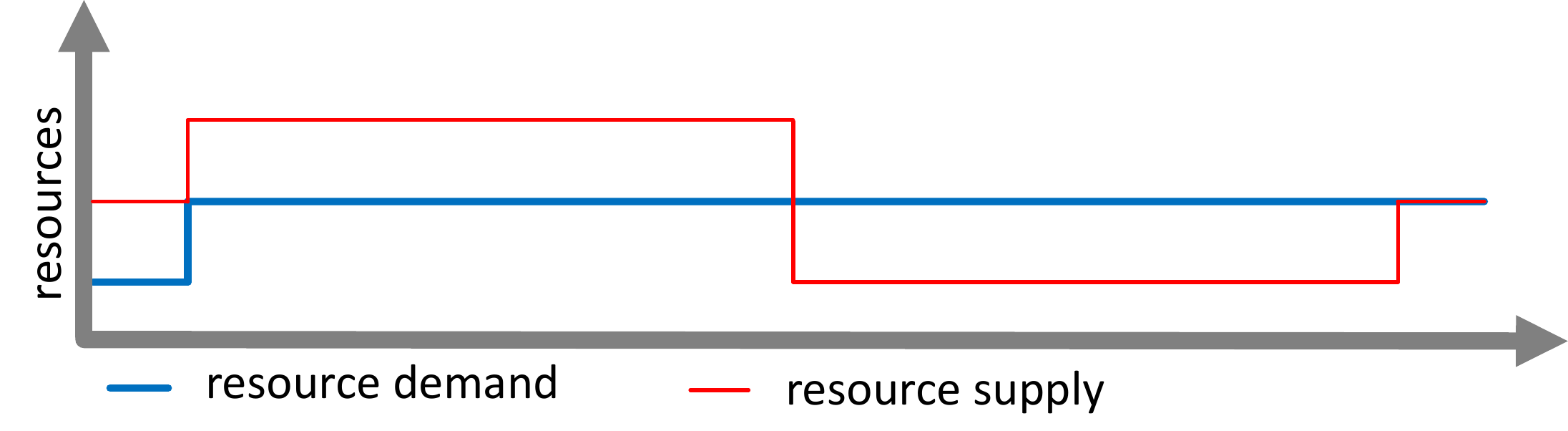}
    	\label{fig:metrics:jitter:low}
    }
    \subfigure[Platform B]{
   \includegraphics[width=0.8\columnwidth]{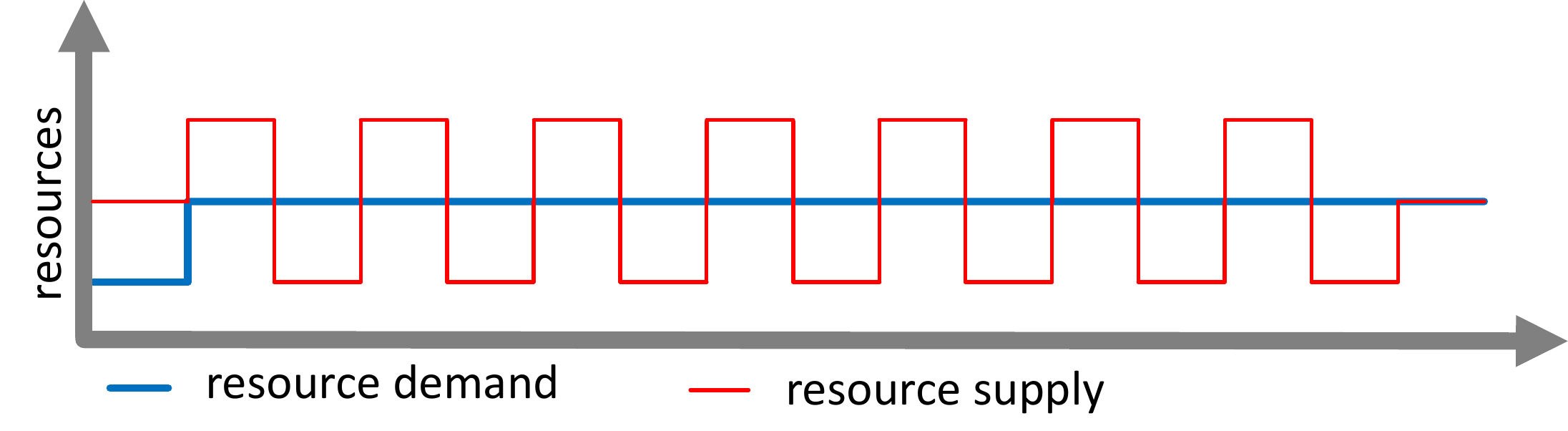}
    	\label{fig:metrics:jitter:high}
    } \caption[Different elastic behaviors produce equal results for $accuracy$
    and $timeshare$]{Platforms with different elastic behaviors that produce
    equal results for $accuracy$ and $timeshare$ metrics}
    \label{fig:metrics:timing:jitter}
\end{figure}

\paragraph{Jitter}
\label{ch:Metrics:sec:Timing:Jitter}
Although the $accuracy$ and $timeshare$ metrics measure important aspects of
elasticity, platforms can still behave very different while producing the
same metric values for $accuracy$ and $timeshare$ metrics. An example is shown
in Figure~\ref{fig:metrics:timing:jitter}.

Both Platforms A and B exhibit the same accuracy metrics and spend
the same amount of time in the under-provisioned respectively over-provisioned
states. However, the behavior of both platforms differs significantly. Platform
B triggers unnecessary resource supply adaptations whereas Platform A
triggers just a few. We propose to capture this with a further
metric called \emph{jitter} to support reasoning for instance-hour-based pricing
models as well as the operators view on minimizing adaptation overheads.

The $jitter$ metric compares the amount of adaptations within the supply curve
$E_S$ with the number of adaptations within the demand curve $E_D$. 
If a platform de-/allocates more than one resource at a time, the adaptations are
counted individually per resource unit.
The difference is normalized with the length of the measurement period $T$:
\begin{flalign*}
  \text{Jitter metric: } jitter~\left[\frac{\#adap.}{min}\right] &= \frac{E_S - E_D}{T}
\end{flalign*}
A negative $jitter$ metric indicates that the platform adapts rather sluggish to
a changed demand. A positive $jitter$ metric means that the platform tends to oscillate like Platforms A
(little) and B (heavily) as in Figure~\ref{fig:metrics:timing:jitter}. High
absolute values of $jitter$ metrics in general indicate that the platform is not
able to react on demand changes appropriately. In contrast to the accuracy and
timeshare metrics, a $jitter$ value of zero is a necessary, but not sufficient
requirement for a perfect elastic system.

\subsection{Metric Aggregation}
\label{ch:Metrics:Compare}
The sections above explained different metrics for capturing core aspects of
elasticity:
\begin{itemize}
	\item $accuracy_U$ and $accuracy_O$ measure average resource amount
	deviations 
	\item $timeshare_U$ and $timeshare_O$ measure ratios of time in 
	under- or over-provisioned states
	\item $jitter$ measures the difference in demand and supply changes 
\end{itemize}
For a more comfortable comparison of platforms, we propose a way to aggregate
the metrics and to build a consistent and fair ranking, similar to the
aggregation and ranking of results in established benchmarks, e.g., SPEC CPU2006\footnote{SPEC CPU2006:
\url{http://www.spec.org/cpu2006/}}.

Our proposed approach to compute an aggregated
\textit{elastic speedup} consists of the
following three steps:
\begin{enumerate}
	\item {Aggregate the accuracy and timeshare sub metrics into a weighted accuracy and a weighted timeshare metric, respectively.}
	\item {Compute elasticity speedups for both of the aggregated metrics using the
	metric values of a baseline platform.}
	\item {Use the geometric mean to aggregate the speedups for $accuracy$ and $timeshare$ to a $elastic$ $speedup$ measure.}
\end{enumerate}
The resulting $elastic$ $speedup$ measure can be used to compare platforms without
having to compare each elasticity metric separately. As a limitation of this
approach, the $jitter$ metric should not be included (as it can become
zero also in realistic cases of imperfect elasticity). Each of the three steps
is explained in the following.
\begin{enumerate}
	\item The $accuracy_U$ and $accuracy_O$ metrics are combined to a
	weighted accuracy metric $accu$\-$racy_{weighted}$:
	\begin{flalign}
		accuracy_{weighted} =& ~w_{acc_U} \cdot accuracy_U + w_{acc_O} \cdot accuracy_O\\
		\text{ with } 
		&w_{acc_U}\text{, } w_{acc_O} \in [0,1],~~w_{acc_U} + w_{acc_O} = 1
	\end{flalign}
	In the same way, the $timeshare_U$ and $timeshare_O$ metrics are combined to a weighted timeshare metric $time$\-$share_{weighted}$:
	\begin{flalign}
		timeshare_{weighted} =& ~w_{ts_U} \cdot timeshare_U + w_{ts_O} \cdot timeshare_O\\
		\text{ with } 
		&w_{ts_U}\text{, } w_{ts_O} \in [0,1],~~w_{ts_U} + w_{ts_O} = 1
	\end{flalign}
	\item Let $x$ be a vector that stores the metric results:
	\begin{flalign}
		x = \left(x_{1},x_{2}\right) = \left(accuracy_{weighted}, timeshare_{weighted}\right)
	\end{flalign}
	For a metric vector $x_{base}$ of a given baseline platform and a metric vector
	$x_k$ of a benchmarked platform $k$, the speedup vector $s_k$ is computed as
	follows:
	\begin{flalign}
		s_k = \left(s_{k_{accuracy}},s_{k_{timeshare}}\right) = \left(\frac{x_{base_1}}{x_{k_1}}, \frac{x_{base_2}}{x_{k_2}}\right)
	\end{flalign}
	\item 
	The elements of $s_k$ are aggregated to an unweighted $elastic~speedup_{{unweighted}_k}$ measure using the geometric mean:
	\begin{flalign}
		elastic~speedup_{{unweighted}_k} = \sqrt{s_{k_{accuracy}}\cdot s_{k_{timeshare}}}
	\end{flalign}
The geometric mean produces consistent rankings, no matter how measurements are
normalized and is the only correct mean for normalized measurements
\cite{Fleming1986}.
Thus, the ranking of the platforms according to $elastic$\-$~speedup_k$ is
consistent, regardless of the chosen baseline platform. 
Furthermore, different preferences concerning the elasticity aspects can be
taken into account by using the weighted geometric mean for computing the
$elastic~speedup_{{weighted}_k}$:
	\begin{flalign}
		elastic~speedup_{{weighted}_k} =& ~{s_{k_{accuracy}}}^{w_{acc}}\cdot{s_{k_{timeshare}}}^{w_{ts}}\\
	\text{ with }
	& w_{acc}\text{, } w_{ts} \in [0,1], ~~w_{acc} + w_{ts} = 1
	\end{flalign}
\end{enumerate}

The following equation summarizes the three steps:

\begin{center}
$elastic~speedup_{{weighted}_k} = $

${\left(\frac{w_{acc_U} \cdot {acc_U}_{base}~+~w_{acc_O} \cdot
{acc_O}_{base}}{{{w_{acc_U} \cdot {acc_U}_{k}}~+~{w_{acc_O} \cdot
{acc_O}_{k}}}}\right)}^{w_{acc}}~\cdot\notag {\left(\frac{w_{ts_U} \cdot
{ts_U}_{base}~+~w_{ts_O}~\cdot~{ts_O}_{base}}{{{w_{ts_U} \cdot
{ts_U}_{k}}~+~{w_{ts_O} \cdot {ts_O}_{k}}}}\right)}^{w_{ts}}$\\

$\text{with }\notag w_{acc_U}\text{, } w_{acc_O}\text{, } w_{ts_U}\text{, }
w_{ts_O},~w_{acc}\text{, } w_{ts} \in [0,1];\notag$ 

$w_{acc_U} + w_{acc_O} = 1;~~w_{ts_U} + w_{ts_O} = 1;~~w_{acc} + w_{ts} =
1\notag $\\
\end{center}

\paragraph*{Elasticity Metric Weights}
\label{ch:Metrics:Compare:Weights}

A single number measuring the elasticity as shown in the equation above can be
adjusted according to the preferences of the target audience by using different weights.
For example, the accuracy weights $w_{acc_U} = 0.2,~~ w_{acc_O} = 0.8$ allow to amplify the
influence of the amount of over-provisioned resources compared to the amount of
under-provisioned resources. A reason for doing so could be that
over-provisioning leads to additional costs, which mainly depend on the amount
of over-provisioned resources. The cost for under-provisioning in contrast does
not depend that much on the amount of under-provisioned resources but more on
how long the platform under-provisions. This observation can be taken into account
by using timeshare weights like: $w_{ts_U} = 0.8,~~ w_{ts_O} = 0.2$. Finally,
when combining the \textit{accuracy} and \textit{timeshare} speed ups, the
weights $w_{acc},~ w_{ts}$ can help to prioritize different elasticity aspects.
Here, weights like $w_{acc} = \frac{1}{3},~~ w_{ts} = \frac{2}{3}$ for example
would double the importance short under- and over-provisioning periods compared
to the importance of small under- or over-provisioning amounts.

\subsection{Elasticity Measurement Approach}

\begin{figure}[!htb]
			\begin{center}
				\includegraphics[width=0.9\columnwidth]{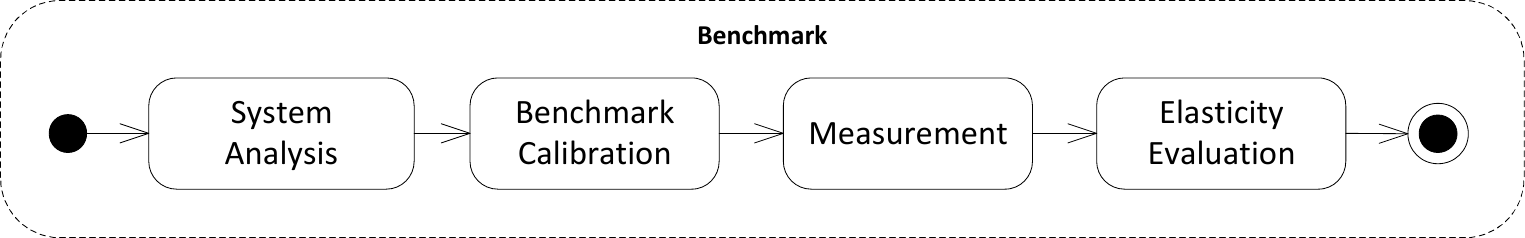}
				\caption{Activity diagram for the benchmark work flow}
				\label{fig:BenchmarkWorkflow}
			\end{center}
		\end{figure}

	This paragraph shortly sketches an elasticity benchmarking concept that we propose as described in \cite{HeKoWeGr_2015_SEAMS_BUNGEE} together with its implementation called
	BUNGEE\footnote{BUNGEE Cloud Elasticity Benchmark:
		\url{http://descartes.tools/bungee}}. 
	Generic and cloud specific benchmark requirements as stated in ~\cite{Huppler2009,Huppler2012} and \cite{Folkerts2012} are considered in this approach.
	Figure~\ref{fig:BenchmarkWorkflow} shows the four main steps in the measurement
	process shortly explained in the following:

	\begin{enumerate}
		\item {\textbf{Platform Analysis: }}
		The benchmark analyzes the system under test (SUT) with respect to the
		performance of its underlying resources and its scaling behavior.
		\item {\textbf{Benchmark Calibration: }}
		The results of the analysis are used to adjust the load intensity profile
		injected on the SUT in a way that it induces the same resource demand on all
		compared platforms.
		\item {\textbf{Measurement: }}
		The load generator exposes the SUT to a varying workload according to the
		adjusted load profile. The benchmark extracts the actual induced resource
		demand and monitors resource supply changes on the SUT.
		\item {\textbf{Elasticity Evaluation: }}
		Elasticity metrics are computed and used to compare the resource demand 
		and resource supply curves with respect to different elasticity aspects.
	\end{enumerate}
	
\begin{figure}[!htb]
		\begin{center}
			\includegraphics[width=1\columnwidth,
			trim=0.5cm 0cm 0cm 0cm,
			clip=true]{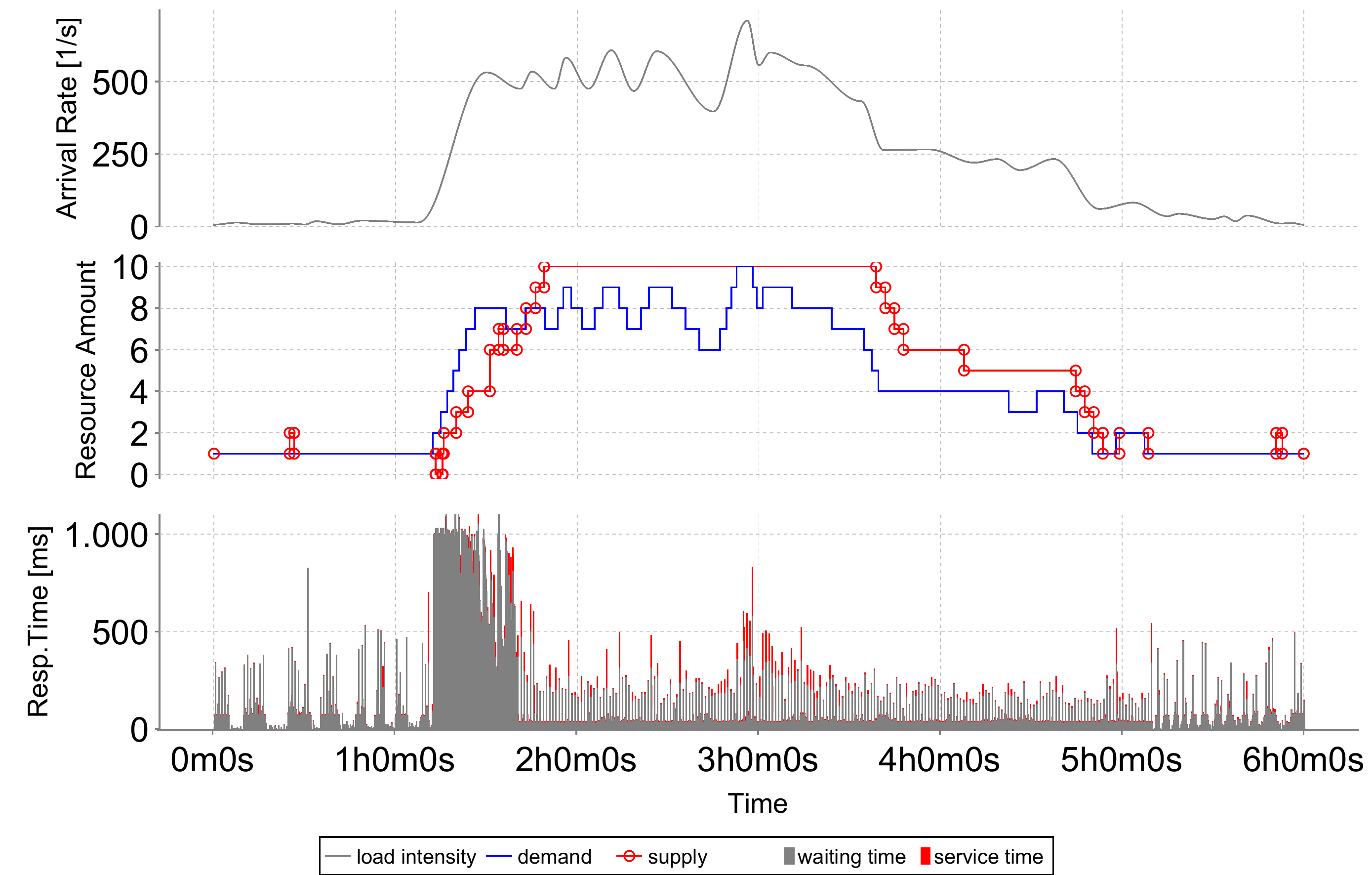}
			\caption{Plot resource demand and supply for an exemplary benchmark run on
				a public cloud}
			\label{fig:eval:6h:aws:ConfigH1}
		\end{center}
	\end{figure}

	The results of an exemplary benchmark run are plotted in
	Figure~\ref{fig:eval:6h:aws:ConfigH1} and the computed elasticity metrics in
	Table~\ref{table:eval:6h:awsAndCS:MetricResults}
	
			\begin{table}[!htb]
				\caption[Metric results for an exemplary benchmark run]{Metric results for an
					exemplary benchmark run}
				\label{table:eval:6h:awsAndCS:MetricResults}
				\centering
				\begin{tabular}{>{\centering}m{1.5cm}*{5}{>{\RaggedLeft}m{1.5cm}}}
					$acc_O$\newline [\#res.] & $acc_U$ \newline[\#res.] & $ts_O$\newline [\%] &
					$ts_U$\newline [\%] &$jitter$\newline$\left[\frac{\#adap.}{min}\right]$\\
					1.053 & 0.180 & 51.9 & 8.1 & -0.033\\
					\vspace{1mm}
				\end{tabular}
			\end{table}

\subsection{Discussion}

This section presents a set of metrics for capturing the accuracy and timing aspects of elastic platforms. Existing cost-based and end-user focused metrics are strongly dependent on the provider's cost model and therefore not seen as independent for valid cross-platform comparisons. Low-level technical elasticity metrics like mean provisioning time leave out the impact of an auto-scaler configuration.
We provide a metric aggregation method based speed-up ratios for relative comparisons and showed how it can be adapted to personal preferences using weights. 
We shortly describe a corresponding elasticity benchmarking methodology that enables cross-platform comparisons even if the performance of underlying resource units differs.

\section{Performance Isolation} \label{sec:PerfIsolation}
In this section, metrics and techniques for quantifying performance isolation
based on current research being presented. Two different methodologies and
several alternative metrics along with appropriate measurement techniques for
quantifying the isolation capabilities of IT systems with help of performance
benchmarks.

\subsection{Goal and Relevance}
Cloud Computing shares resources among several customers on various layers like
IaaS, PaaS or SaaS. The isolating of cloud customers with regards to performance
is one of the major challenges to achieve reliable performance.

The allocation of hardware resources is handled by the lower levels (e.g.,
infrastructure level) in the stack. Therefore, performance isolation is a bigger
challenge in the upper levels (e.g., platform and software) as they
intentionally have no direct resource control. Nevertheless, also on the
infrastructure level we can observe significant influence between virtual
machines as described in \cite{HuQuHaKo2011-CLOSER-ModelVirtOverhead}.

Performance isolation is an important aspect for various stakeholders. When a
developer or architect has to develop a mechanism to ensure performance
isolation between customers they need to validate the effectiveness of their
approach to ensure the quality of the product. Furthermore, to improve an
existing mechanism they need an isolation metric to compare different variants
of the solution. When a system owner has to decide for one particular deployment
in a virtual environment not only traditional questions like the separation of
components on various hosts are of importance but also how the deployments
non-functional runtime properties will be influenced. 
the hypervisor with regards to resource allocation mechanism have to be
considered.
concerns might be important.

\subsection{Foundations for the Metrics}
The subsequently presented Metrics follow the ideas presented in \cite{Krebs2012, KrMoKo2013-SciCo-MetricsAndTechniquesForPerformanceIsolation}.
\paragraph{Definition of Performance Isolation}
Performance concerns in cloud environments are a serious obstacle for consumers.
To avoid distrust, it is necessary to ensure a fair behavior. This means,
Customers working within their assigned quota should not suffer from customers
exceeding their quotas. Quota refers to the amount of workload a customer is
allowed to execute one metric might be the request rate.

\paragraph{Performance Isolation}
A system is performance-isolated, if for customers working within their quotas
the performance is not affected when other customers exceed their quotas. A
decreasing performance for the customers exceeding their quotas is accepted. It
is possible to relate the definition to SLAs: A decreased performance for the
customers working within their quotas is acceptable as long as it is within
their SLAs.

\paragraph{Basic Idea}
The metrics defined may be applied to quantify the isolation of any measurable
QoS-related system property in any system shared between different entities. Of
course, the actual type of workload and QoS must be selected according to the
scenario under investigation.

\begin{figure}[htb]
\centering
\includegraphics[width=0.7\columnwidth]{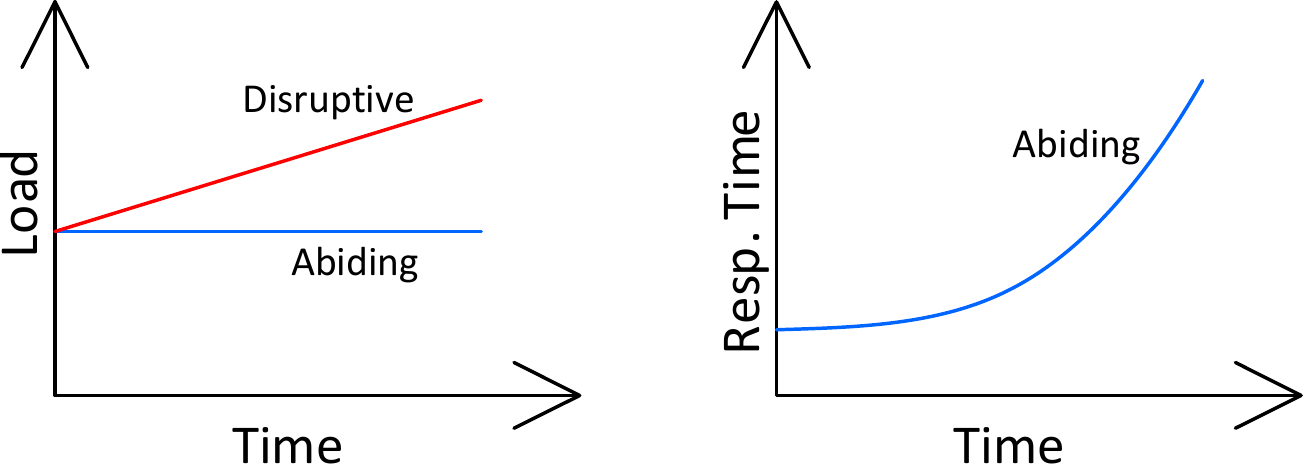}
\caption{Influence of the disruptive tenant onto the abiding.}
\label{fig:ExplainedIncrease}
\end{figure}

The metrics distinguish between groups of \emph{disruptive} and \emph{abiding}
customers. The latter work within their given quota (e.g., defined number of
requests/s) the former exceed their quota. Isolation metrics are based on the
influence of the disruptive customers on the abiding customers. Thus we have two
groups and observe the performance of one group as a function of the workload of
the other group (cf.\ Figure~\ref{fig:ExplainedIncrease}).

To provide a level playing field for comparisons, a description of a workload profile $W$ is required together with the metrics. Although, the concrete definition of the workload $W$ is a case to case decision and usually defined to answer a dedicated question, we share some general thoughts concerning this issue. Multi-tenant cloud applications (MTAs) are operated at rather high utilization for economic reasons. 
Another important reason to run the test system under high utilization, is the goal to evaluate performance isolation aspects. 
In a system with low utilization of the resources, the increased workload of one tenant has a low impact upon the performance of the others, as no bottleneck exists. 
Another aspect is related to existing guarantees. 
If the provider wants to maintain a certain QoS, it is possible to configure the overall reference workload in a way the average systems QoS is close to this value. 
In this case, a small increase of workload at the disruptive tenant, immediately results in violations for the abiding ones, in case of a weak isolation. 
In case no SLA based guarantee exists, and the bottleneck resource is unknown, a measurement to identify the system’s maximum throughput, by increasing all tenants workloads in parallel is feasible. 
To increase the speed finding this point, a binary like search can be used. 
Usually this workload is accompanied with the highest possible utilization of the bottleneck resource. 
This means, that increasing the workload in a non-isolated system, will immediately result in less performance for all other tenants. Consequently, it would immediately cause guarantee violations. Another argument is, that an isolation mechanisms should intervene latest at this point. The key findings for the workload $W$ are summarized as follows:
(I)~High load/utilization preferable, (II) QoS observed should be close to the guarantee, and (III) the systems maximum throughput can be used as an indicator.

\subsection{Metrics based on QoS impact}
For the definition of the metrics, a set of symbols are defined in Table
\ref{tab:Symbols}.

\begin{table}[ht]

\begin{center}
\begin{tabular}{|p{1.1cm}|p{12cm}|}
  \hline
  Symbol &  Meaning\\
  	\hline\hline
  		$t$ & A customer in the system.\\ 
  	\hline
		$D$ & Set of disruptive customers exceeding their quotas (e.g., contains
		customers inducing more than the allowed requests per second). $|D|>0$ \\
	\hline
		$A$ & Set of abiding customers not exceeding their quotas (e.g., contains
		customers inducing less than the allowed requests per second).$|A|>0$\\
	\hline
		$w_{t}$ & Workload caused by customer $t$ represented as numeric value $\in \mathbb{R}_{0}^{+}$. The workload is considered to increase with higher values (e.g., request rate and job size). $w_{t} \in W$\\
	\hline
		$W$ & The total system workload as a set of the workloads induced by all
		individual customers. Thus, the load of the disruptive and abiding ones. \\
	\hline
		$z_{t}(W)$ & A numeric value describing the QoS provided to customer $t$. The
		individual QoS a customer observes depends on the composed workload of all
		customer $W$. We consider QoS metrics where lower values of $z_{t}(W)$ correspond to better qualities (e.g., response time) and $z_{t}(W) \in \mathbb{R}_{0}^{+} $ \\
	\hline
		$I$ & The degree of isolation provided by the system. An index is added to distinguish different types of isolation metrics. The various indices are introduced later. 
		\\
  \hline
\end{tabular}
\end{center}
	\caption{Overview of variables and symbols}
	\label{tab:Symbols} 
\end{table}

These metrics depend on at least two measurements. First, the observed QoS
results for every $t \in A$ at a reference workload $W_{ref}$. Second, the
results for every $t \in A$ at a workload $W_{disr}$ when a subset of the
customers have increased their load to challenge the system's isolation
mechanisms. As previously defined $W_{ref}$ and $W_{disr}$ are composed of the
workload of the same set of customers which is the union of $A$ and $D$.  At
$W_{disr}$ the workload of the disruptive customers is increased.

We consider the relative difference of the QoS ($\Delta z_A$) for abiding
customers at the reference workload compared to the disruptive workload.
\begin{equation}
\Delta z_{A}=
\frac{\sum\limits_{t \in A}[ z_{t}(W_{disr})-z_{t}(W_{ref})]} 
{\sum\limits_{t \in A} z_{t}(W_{ref})}
\end{equation}
Additionally, we consider the relative difference of the load induced by the
two workloads.
\begin{equation}
\Delta w = \frac{\sum\limits_{w_{t} \in W_{disr}}^{}w_{t} - \sum\limits_{w_{t} \in W_{ref}}^{}w_{t}}
{\sum\limits_{w_{t} \in W_{ref}}^{}w_{t}}
\end{equation}
Based on these two differences the influence of the increased workload on the
QoS of the abiding tenants is expressed as follows.
\begin{equation}
I_{QoS}=\frac{\Delta z_{A}}{\Delta w}
\end{equation}
A low value of this metric represents a good isolation as the difference of the
QoS in relation to the increased workload is low. Accordingly, a high value of
the metric expresses a bad isolation of the system.

Another metric is an enhancement of the previous one, considering the arithmetic
mean of $I_{QoS}$ for $m$ disruptive workloads. Whereby the disruptive customers
increase their workload equidistant within a lower and upper bound.
\begin{equation}
\label{eq:average}
I_{avg}=\frac{\sum\limits_{i=1}^{m}I_{QoS_{m}}}{m}
\end{equation}

\subsection{Workload Ratios}
\begin{figure}[htb]
\centering
\includegraphics[width=0.7\columnwidth]{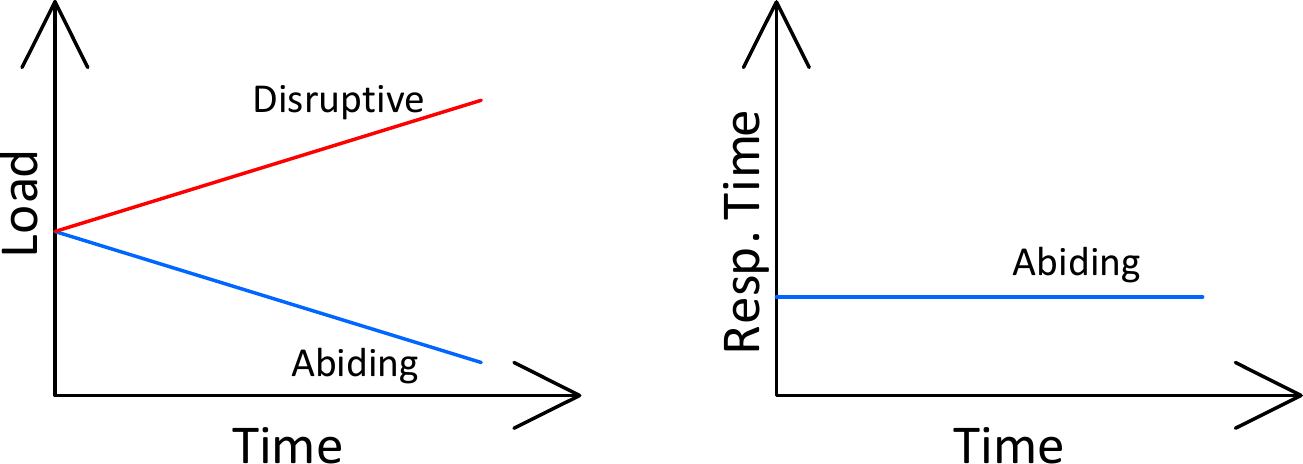}
\caption{Workload Adaption to maintain QoS.}
\label{fig:ExplainedNoIncrease}
\end{figure}

The following metrics are not directly associated with the QoS impact resulting
from an increased workload of disruptive customers as it was depicted in Figure~\ref{fig:ExplainedIncrease}. The idea is to compensate
the increased workload of disruptive customers and try to keep the QoS for the
abiding ones constant by decreasing the workload of the abiding ones (cf.\ Figure~\ref{fig:ExplainedNoIncrease}).

Naturally, this is only possible with the support of the abiding customers and
such a behavior does not reflect productive systems. Thus, these metrics are
planned to be applied in benchmarks with artificial workloads where a load
driver emulates the customers and can be enhanced to follow this
behavior. 

Assume one starts measuring the isolation behavior of a non-isolated system by
continually increasing the disruptive workload $W_{d}$. One would expect to
observe a decrease of $z_{t}(W)$ for all customers. In such a situation,
$z_{t}(W)$ would remain unaffected if the workload of the abiding customers
$W_{a}$ is decreased accordingly to compensate for the increase in the
disruptive workload. Following this idea, plotting $W_{a}$ as a result of
$W_{d}$ describes a Pareto optimum of the systems total workload with regards to
constant QoS.

\begin{figure}[htb]
\centering
\includegraphics[width=0.7\columnwidth]{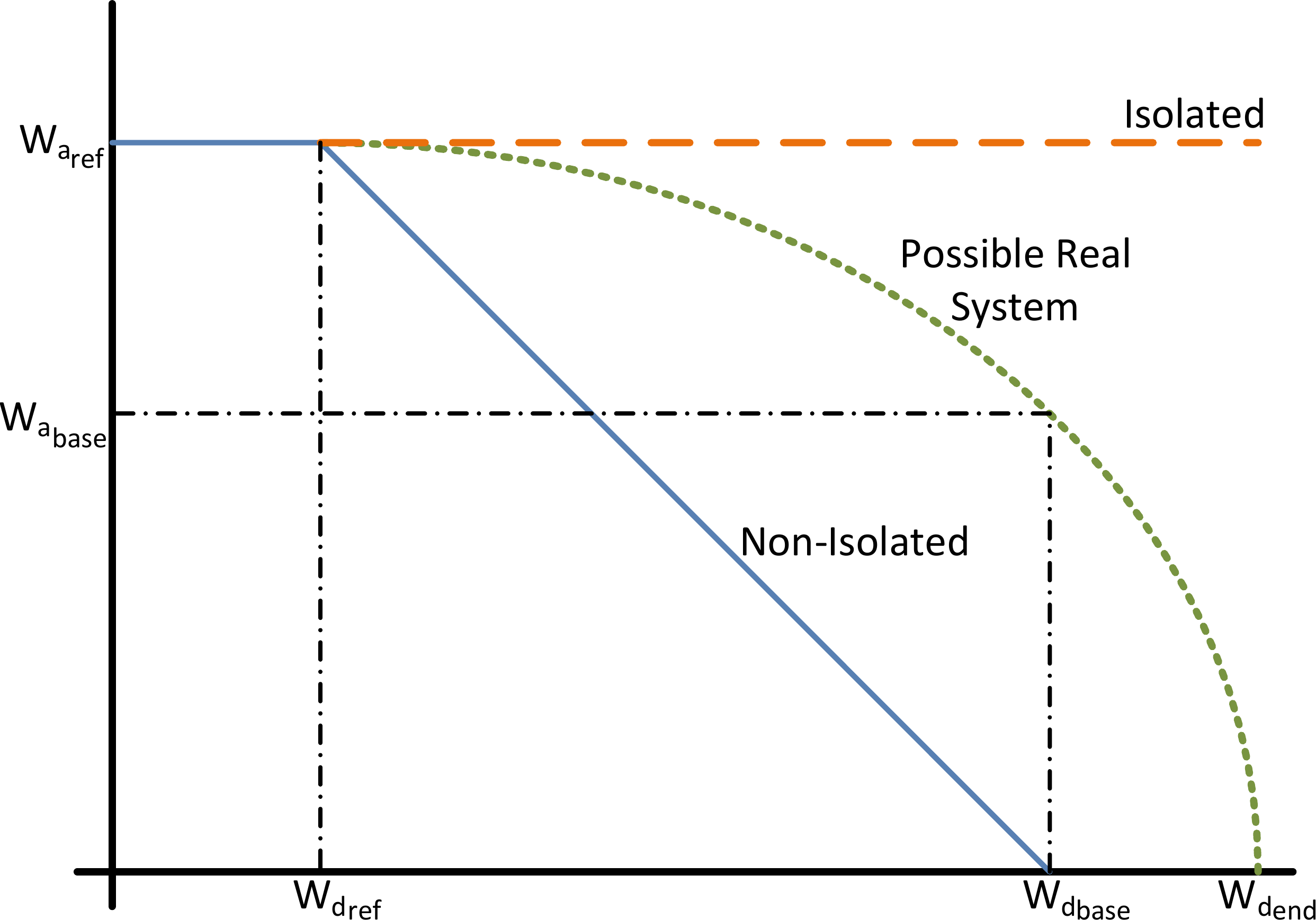}
\caption{Fictitious isolation curve including upper and lower bounds.}
\label{fig:IsolationBounds}
\end{figure}

In Figure \ref{fig:IsolationBounds}, the x-axis shows the amount of workload
$W_{d}$ caused by the disruptive tenants, whereas the y-axis shows the amount of
the workload $W_{a}$ caused by the abiding tenants. The blue/solid line shows
how $W_{a}$ has to decrease to maintain the same QoS as in the beginning. In a
non-isolated system this function proportionally decrease linear. For every
additional amount added to the disruptive load one has to remove the same amount
at the abiding load, because in a non-isolated system the various workload
groups would behave if they were one.
In a perfectly isolated system the increased $W_{d}$ has no influence on
$z_{t}(W)$ for all $t \in A$. Thus, $W_{a}$ would be constant in this case as
shown with the red/dashed line in the figure. The red line and the blue line
provide exact upper and lower bounds, corresponding to a perfectly isolated and
a non-isolated system, respectively. Figure \ref{fig:IsolationBounds} shows some
important points referenced later and defined in Table
\ref{table:ReferencePoints}.

\begin{table}[ht]
	\begin{center}

\begin{tabular}{|p{1.1cm}|p{12cm}|}

  \hline
  		Symbol & Definition\\
  	\hline \hline
		$W_{d}$ & The total workload induced by the disruptive customers: \newline
$
W_{d}=
\sum\limits_{t \in D} w_{t}
$
\\
	\hline
		$W_{d_{base}}$ & The level of the disruptive workload at which the abiding
		workload in a non-isolated system is decreased to 0 due to SLA violations.\\
	\hline
		$W_{d_{end}}$ & The level of the disruptive workload at which the abiding
		workload must decreased to 0 in the system under test\\
	\hline
		$W_{d_{ref}}$ & The value of the disruptive workload at the reference point
		in the system under test. This is the point to which the degree of isolation
		is quantified. It is defined as the disruptive workload, at which in a
		non-isolated system the abiding workload begins to decrease.\\
	\hline
		$W_{a}$ & The total workload induced by the abiding customers: \newline
$
W_{a}=
\sum\limits_{t \in A} w_{t}
$
\\
	\hline
		$W_{a_{ref}}$& The value of the abiding workload at the reference point
		$W_{d_{ref}}$ in the system under test. $ W_{a_{ref}} = W_{d_{base}} -
		W_{d_{ref}}$ \\
  \hline
		$W_{a_{base}}$ & The value of the abiding workload corresponding to
		$W_{d_{base}}$ in the system under test. \\
	\hline
\end{tabular}
\end{center}
\caption{Overview and definition of relevant points.}
\label{table:ReferencePoints}
\end{table}


Based on this approach, several metrics presented in the following.
As discussed before, the workload scenarios play an important role, and thus it
may be necessary to consider multiple different workload scenarios and average
over them.

\paragraph{Significant Points}
The significant points marked in Figure \ref{fig:IsolationBounds} provide
several ways to define an isolation metric by themselves. $I_{end}$ is a metric
derived by the point at which the workloads of abiding customers have to be
decreased to 0 to compensate for the disruptive workload. The metric sets
$W_{d_{end}}$ and $W_{a_{ref}}$ in relation. Due to the discussed relationship
of the workloads in a non-isolated system and the definition of the various
points based on the behavior of such a system the condition $ W_{a_{ref}} =
W_{d_{base}} - W_{d_{ref}}$ holds. We leverage this relation to simplify our
formulas. With Figure \ref{fig:IsolationBounds} in mind, $I_{end}$ is defined as
follows:
\begin{equation}
I_{end}=\frac{W_{d_{end}}-W_{d_{base}}}{W_{a_{ref}}}
\end{equation}
Another approach uses $W_{a_{base}}$ as a reference. Setting this value and
$W_{a_{ref}}$ in relation results in an isolation metric having a value between
$[0,1]$. The formula for metric $I_{base}$ is below:
\begin{equation}
I_{base}=\frac{W_{a_{base}}}{W_{a_{ref}}}
\end{equation}

For systems that exhibit a linear degradation of abiding workload, we could also
define isolation metrics based on the angle between the observed abiding
workloads line segment and the line segment which represents a non-isolated
system. However, linear behavior typically cannot be assumed.

\paragraph{Integral Metrics}
These metrics are based on the area under the curve derived for
the measured system $A_{measured}$ set in relation to the area under the curve
corresponding to a non-isolated system $A_{nonIsolated}$. The area covered by
the curve for a non-isolated system is calculated as $W_{a_{ref}}^2/2$.

The first metric $I_{intBase}$ represents the isolation as the ratio of
$A_{measured}$ and $A_{nonIsolated}$ within the interval 
$[W_{d_{ref}},W_{d_{base}}]$. Let $f_{m}:W_{d}\rightarrow W_{a}$ be a
function which returns the residual workload for the abiding customers based on
the workload of the disruptive customers. We then define the metric
$I_{intBase}$ as follows:
\begin{equation}
I_{intBase}=\frac{\left(\int\limits_{W_{d_{ref}}}^{W_{d_{base}}}f_m(W_{d})
\textit{d} W_{d}\right) - W_{a_{ref}}^2/2} {W_{a_{ref}}^2/2}
\end{equation} 
$I_{intBase}$ has a value of 0 in cases the system is not isolated and a value
of 1 if the system is perfectly isolated within the interval $[W_{d_{ref}},
W_{d_{base}}]$. 

The following metric $I_{intFree}$
allows to use any predefined artificial upper bound $p_{end}$ which represents
the highest value of $W_d$ that was measured in the system under test. We
define the metric as follows:
\begin{equation}
I_{intFree}=\frac{\left(\int\limits_{W_{d_{ref}}}^{p_{end}}f_m(W_{d})
\textit{d} W_{d}\right) - W_{a_{ref}}^2/2} {W_{a_{ref}} \cdot
(p_{end}-W_{d_{ref}})-W_{a_{ref}}^2/2}
\end{equation} 
This metric quantifies the degree of isolation provided by the system for a
specified maximum level of injected disruptive workload $p_{end}$. A value of
1 represents a perfect isolation and a value of 0 a non-isolated system.

\subsection{Measurement Methodology}
\paragraph{System Setup}
All isolation metrics are based on the observation of QoS metrics. In Cloud
Systems an increasing workload may lead to the allocation of additional
resources to keep the observed QoS constant. This behavior belongs to
Elasticity \ref{sec:Elasticity}. To really measure the systems isolation
capabilities one has to ensure a constant amount of resources and hardware
setup.

\paragraph{Selection of QoS and Workload}
For a concrete measurement of the particular isolation for one quality metric
and workload definition of a system, one has to select an appropriate metric.
These metrics have to follow the guidelines discussed at the metrics. Especially
one has to consider, that the chosen metric for the workload has to be
equivalent for the two groups of users.
 
\paragraph{Workload Profile}
For the definition of the workload profile used for measurement we have to
consider especially the reference workload profile. The definitions of the upper
bounds of the measurements are strongly scenario dependent but might expressed
as multiple of the reference workload to make it comparable between various
systems. It is conceivable that the observed impact of increasing workload onto
the QoS (especially for performance metrics) is rather low when the system is
low utilized. Therefore the reference workload should relate to the maximum
throughput a system could achieve. As of this point an increasing load of the
disruptive customers has a high impact onto the systems performance.

\subsection{Discussion}

\paragraph{QoS Impact} 
This metric helps system owners to manage their systems, because it indicates
the influence of disruptive workload onto the QoS they provide, which is
important for capacity planning. QoS-based metrics can prove that a system is
perfectly isolated, however they fail in ranking a systems isolation
capabilities into the range between isolated and non-isolated. A single
$I_{QoS}$ metric can be derived with only two measurements to obtain evidence
for one point of increased workload. However, to obtain some more detailed
information concerning the systems isolation more measurements are required.
Therefore, $I_{avg}$ describes the average isolation value within the upper and
lower bound of interest.

\paragraph{Significant Points} 
The metric $I_{end}$ might not be feasible to quantify isolation in well
isolated systems. Furthermore, it is not possible to directly deduce relevant
system behavior like response times. If the metric is given, it could help to
compare two systems regarding the maximum disruptive load they can handle. To
determine $I_{end}$, more measurements as for QoS-based metrics are required.

$I_{base}$ orders a system within the range of isolated and non-isolated systems
for one specific point in the diagram. Nevertheless, it does not provide
information about the behavior of the system before that point. It is limited
to comparing the isolation behavior of the systems at one selected load level
and it is inadequate to derive direct QoS-related values. The usefulness of this
metric appears to be of limited value in contrast to the integral methods. One
advantage is the evidence at a well-defined and reliable point with only two
measurements.

Both metrics have some drawbacks resulting from the fact that they do not take
into account the curve progression. This means, that in a system which behaves
linearly until a short distance from $W_{d_{base}}$ and then suddenly drops to
$W_a=0$, both metrics would have the same value as in the case of a completely
non-isolated system which is obviously unfair in this case. Moreover, a well
isolated system might require a very high disruptive workload before $W_a$ drops
to $0$ making it hard to measure the metric in an experimental environment.
$I_{base}$ has some further disadvantages given that it is only representative
for the behavior of the system within the range of $W_{d_{ref}}$ and
$W_{d_{base}}$. Given that the metric does not reflect what happens after
$W_{d_{base}}$, it may lead to misleading results for well isolated systems
whose respective $W_{d_{end}}$ points might differ significantly.

\paragraph{Integral Metrics}
$I_{intBase}$ and $I_{intFree}$ are widely comparable metrics. $I_{intBase}$ has
the advantage to be measured at a predefined point. For $I_{intFree}$, the
endpoint of the interval must be considered as well to have an expressive
metric. Both metrics provide good evidence of the isolation within the
considered interval, ordered between the magnitudes of isolated and non-isolated
systems. They lack in providing information concerning the degree of SLA
violation. For example, the SLA violation could be very low and acceptable or
critically high in each iteration when we reduce $W_a$. However, in both cases,
the results of the metrics are similar. This limits the value of $I_{intBase}$
and $I_{intFree}$ for system owners/providers. However, for comparison of
systems and analyzing their behavior, the metrics are very useful and can be
exploited by developers or researchers. These metrics do not reflect real
systems and consequently it is very unlikely to use them outside benchmarking
scenarios.
\section{Availability}\label{sec:Availability}

\subsection{Goal and Relevance}

The goal of this section is to identify critical aspects of measuring
Availability of Cloud services, existing metrics and their relation to the de
facto standards and conditions that apply to the public Cloud market. Relevance
to Cloud environments is high, given that Availability is one of the strong
arguments for using Cloud services, together with the elastic nature of the
resources and adaptable utilization. Furthermore, availability is a key performance indicator (KPI)
included in the majority of public Cloud Service Level Agreements (SLAs).

The final goal is to identify metrics that can be used in various aspects such
as provider and service comparisons or potential incorporation into Trust and
Reputation mechanisms. Furthermore we intend to highlight aspects that are
needed for an Availability benchmark, which in contrast to most cases of
benchmarking, it does not refer to the creation of an elementary computational
pattern that may be created in order to measure a system's performance but
mainly to a daemon-like monitoring client for auditing provider SLAs and
extracting the relevant statistics. For this case critical requirements are
investigated.

\subsection{Prerequisites and Context}
\label{aval_12}

Service Level Agreements are contracts between a service provider and a
potential client, that govern or guarantee a specific aspect of the offered
service. Relevant metrics may include availability, performance aspects, QoS
etc. While in research efforts more advanced aspects have been investigated
(e.g. direct SLA negotiaton and runtime renegotiation as in \cite{5476775}, real-time
QoS aspects as in \cite{5984009} etc.), in main public Cloud environments they appear
as static agreements as in \cite{SLA, SLA2, SLA3}, prepared in advance by the
respective provider, and not available for adaptation to each specific user.
Thus a Cloud user by definition accepts these terms when using the specific
provider's services, accepting de facto the terms and conditions offered.

\subsection{Relevant Metrics Definition}

\paragraph{Operational Availability}

For real time monitoring information regarding services availability, one very
interesting approach is \cite{CloudSleuth}. CloudSleuth has an
extensive network of deployed applications in numerous providers and locations
and continuously monitors their performance with regard to response times and
availability metrics. CloudSleuth's measurement way is not adapted to each
provider's SLA definition, so it cannot be used to claim compensation but it
relates mainly to the definition of operational availability. Furthermore, it
checks the response of a web server (status 200 return type on a GET request).
Thus it cannot distinguish between a failure due to an unavailable VM (case of
provider liability) or due to an application server crash (customer liability in
case of IaaS deployment, provider liability in case of PaaS) or pure application
fault (customer liability). On the other hand, it follows a normal availability
definition that makes it feasible to compare services from different providers,
a process which cannot be performed while following their specific SLA
definitions, since they have differences in the way they define availability.
\begin{equation}
Availability=\frac{TotalSamples-UnavailableSamples}{TotalSamples} 
\end{equation}
\begin{center}CloudSleuth Formula for Availability
\end{center}

In \cite{Chandler2012}, availability is defined in relation to the Mean Time
Between Failures metric, so that the metric avoids cases where small uptime is hidden from the
fact of very low downtime. While this is reasonable, commercial Cloud providers
tend to consider as uptime the entire month duration (as mentioned in the
following chapters) rather than only the actual uptime the services were used by
the end user.  A very thorough monitoring tools analysis is conducted in
\cite{Aceto20132093}.
While numerous tools exist for focusing on service availability (and other non
functional properties),  it is questionable whether the way they are calculating
the specific metrics is compliant to the relevant definitions in commercial
public Clouds and their respective SLAs. In \cite{6319167}, an investigation of
new metrics and relevant measurement tools expands across different areas such as network,
computation, memory and storage. However, availability is again not considered
against the actual SLA definitions.

\paragraph{De Facto Industrial SLAs Examination}

As mentioned in Section \ref{aval_12}, commercial providers follow a
considerably different approach, defining by themselves the metric of
availability that is considered in their respective SLAs. In many cases this
implies a different calculation based on the same samples and a set of
preconditions that are necessary for a specific service deployment to be
included under the offered SLA. In order to investigate the ability to provide
an abstraction layer for this type of services and thus create a generic and
abstracted benchmarking client, one needs to identify common concepts in the way
the SLAs are defined for the various providers (we considered Amazon EC2, Google
Compute and Windows Azure). The first and foremost point of attention is the
generic definition of availability, which is included in Equation 23.
\begin{equation}
\begin{array}{l} {Availability=\frac{TotalMinutesOfMonth-DowntimeMinutes}{TotalMinutesOfMonth} =} \\ {\frac{TotalMinutesOfMonth-\sum DownTimePeriods_{i}  }{TotalMinutesOfMonth} } \end{array}
\end{equation}
\begin{center}Provider based definition of
availability\end{center}

This may seem the same as the one defined in Equation 22, however the providers
typically define the downtime period as a step function (x in minutes):\\

\begin{equation}
\end{equation}

   \hspace{3.5cm}$0$ if $x$<$minimum$ (Google: 5, Amazon: 1, Azure: 1)[min] 
   
\noindent $DownTimePeriod(x)=\{$

   \hspace{3.5cm}$x$ if $x>minimum$\\

Furthermore, one other key point is the existence of preconditions needed for an
SLA to be applicable for a specific group of services used (e.g. need to have
multiple VMs deployed in different availability zones). There are variations
between provider definitions with regard to this aspect, with relation to where
the VMs are deployed or if further action is necessary after the realization
that a virtual resource is unavailable (e.g. restart of the resource). In a
nutshell, the similarities and differences of the three examined providers are the following:

\begin{enumerate}
\item Common concepts

\begin{enumerate}
\item  Quantum of downtime period: providers do not accept that a downtime
period is valid, unless it is higher than a specific quantum of time.
\item  Discount formats as compensation for downtime with relation to the
monthly bill.
\item  Calculation cycle of 1 month.
\item  More than one VM to accept SLA applicability.
\item  VM distribution across at least 2 availability zones.
\item  Simultaneous unavailability for all deployed VMs for an overall sample to
be considered as unavailable.
\end{enumerate}
\item  Differences

\begin{enumerate}
\item  Quantum size. This is considered minor since the format is the same.
\item  Number of discount levels. The three providers offer different discounts
for various levels of deviation from the availability goal. This is considered
minor since the format is the same.
\item  For the Azure Compute case, it seems that more than one instances for the
same template ID must be deployed. However, it seems also that the overall time
interval refers only to the time a VM was actually active.
\item  Restart from template is necessary in Google App Engine before validating an unavailable sample
\end{enumerate}
\end{enumerate}

\subsection{Abstracted and Comparable Metrics}

Given the differences in the aforementioned availability definitions, it is not
feasible to directly compare provider-defined availability metrics between
providers, since these differ in definition. For this reason, it is more
meaningful to either follow a more generic direct definition as done in
\cite{CloudSleuth} or abstract to a more generic concept which is the SLA
adherence level. This can be defined as the ratio of violated SLAs over the
overall examined SLAs. Since SLA period is set to monthly cycles, this may be
the minimum granularity of observation.

\paragraph{SLA Adherence Levels}
 
\begin{equation}
SLA\_ Adherence=\frac{violatedSLAs}{overallObservedSLAs}
\end{equation}
\begin{center} SLA Adherence Metric\end{center}

Special attention must be given for cases that sampling is not continuous,
indicating that the client did not have running services for a given period,
applicable for an SLA. These cases must be removed from such a ratio,
especially for the cases that no violations are examined in the limited
sampling period, given that no actual testing has been performed. If on the
other hand even for a limited period a violation is observed, then this may be
included. Furthermore, SLA adherence may be grouped according to the complexity
of the examined service, as mentioned in more detail in Section 1.5.2.

\paragraph{SLA Strictness Levels}

Besides SLA adherence, other metrics may be defined in order to depict the
strictness of an SLA. As a prerequisite, we assume that the metric must follow a
``higher is stricter'' approach. Stricter implies that it is more difficult for
a provider to maintain this SLA. In order to define such a metric initially one
needs to identify what are the critical factors that may affect strictness
levels. Factors should be normalized to a specific interval (e.g. 0 to 1) and
appropriate levels for them may be defined. Indicative factors may include:

\begin{enumerate}
\item  Size of the minimum continuous downtime period (Quantum of downtime
period \textit{q}). A higher size means that the SLA is more relaxed, giving the
ability to the provider to hide outages if they are less than the defined
interval. The effect of such a factor may be of a linear fashion (e.g. 1-q).
Necessary edges of the original interval (before the normalization process) may
be defined per case, based e.g. on existing examples of SLAs.

\item  Ability to use the running time of the services and not the overall
monthly time, denoted by a Boolean variable \textit{t}. This would be stricter
in the sense that we are not considering the time the service is not running as
available. The effect of such a factor may be of a Boolean fashion (0 false, 1 true)

\item  Percentage of availability that is guaranteed. Again this may be of a
linear fashion, with logical intervals defined by the examined SLAs.

\item  Existence of performance metrics (e.g. response timing constraints).
This may be a boolean feature \textit{x}, however its values may be set to
higher levels (0 or 5). The importance of this will be explained briefly.
\end{enumerate}

The added value of such a metric may be in the case we have to deploy
applications with potentially different characteristics and requirements (as one
would expect). For example, having \textit{soft real-time applications} would
imply that we definitely need to have feature 4. Other less demanding
applications may be accommodated by services whose SLAs are less strict. Thus
suitable \textbf{value intervals} may be adjusted for each feature. If we use a
value of 5 for the true case of feature 4, and all the other features are linked
in such a manner that their accumulative score is not higher than 5, then by
indicating a necessary strictness level of 5 implies on a numerical level that
feature 4 needs definitely to be existent in the selected Cloud service.

Depending on the application types and their requirements and based on
the metric definition, one can define categories of strictness based on the
metric values that correspond to according levels (e.g. medium strictness needs
a score from 2 to 3 etc.). It is evident that such a metric is based only on the
SLA analysis and is static, if the SLA definition is not changed. The indicative
formula for the case of equal importance to all parameters appears in Equation 25.

\begin{equation}
\begin{array}{l} {S=t+(1-s_{1} q)+s_{2} p+x{\rm \; where}} \\ {s_{i} :{\rm
normalization\; factor\; for\; the\; continuous\; variables\; so\; that\;
(s}_{1} {\rm *q)}\in [0,1]{\rm \; and\; (s}_{2} {\rm *p)}\in [0,1]} \\ {t\in \{ 0,1\} ,{\rm \; }x\in \{ 0,1\} } \end{array}
\end{equation}
\begin{center}SLA Strictness definition formula\end{center}

For the normalization intervals, for \textit{p} we have used 99\% and 100\% as
the edges, given that these were the ranges encountered in the examined SLAs.
For \textit{q} we have used 0 and 10 minutes as the edges. 0 indicates the case
where no minimum interval is defined (thus abiding by the formula in Equation
22) and 10 the maximum interval in examined Compute level SLAs. However there
are larger intervals (e.g. 60 minutes) in terms of other layer SLAs (Azure
Storage). The limit to 60 has been tried out in the q' case that is included in
Table 1, along with the example of the other factors and the overall values of
the SLA strictness metrics in the 3 examined public SLAs.
\newpage
\begin{center}\textbf{Table 1: Indicative application of the SLA Strictness
metric on existing public Cloud SLAs}
\vspace{0.5cm}
\begin{small}
\begin{tabular}{|c|c|c|c|c|c|c|c|}
\hline \textbf{Provider/Service} & \textbf{\textit{t}} & \textbf{\textit{q}} & \textbf{\textit{q'}} & \textbf{\textit{p}} & \textbf{\textit{x}} & \textbf{\textit{S}} & \textbf{\textit{S'}} \\ \hline 
Google Compute & 0 & 5 (norm:0.5 ) & 5 (norm: 0.0833) & 99.95 (norm:0.5) & 0 & 1 & 1.4167 \\ \hline 
Amazon EC2 & 0 & 1(norm: 0.1) & 1(norm: 0.0167) & 99.95 (norm:0.5) & 0 & 1.4 & 1.4833 \\ \hline 
Microsoft Azure & 1 & 1(norm: 0.1) & 1(norm: 0.0167) & 99.95 (norm:0.5) & 0 & 2.4 & 2.4833 \\ \hline 
\end{tabular}
\end{small}
\end{center}

An example of \textit{x} not being 0 would be the Azure Storage SLA, where
unavailability is also determined by response time limits to a variety of service calls.

\subsection{Measurement Methodology}

In order to create an Availability Benchmark for the aforementioned SLAs, the
following requirements/steps in the methodology must be undertaken:

\begin{enumerate}
\item  Complete alignment to each provider definition in order to achieve
\textbf{Non-repudiation}, including the following:

\begin{enumerate}
\item  Necessary \textbf{preconditions} checking in terms of number and type
\item  Availability \textbf{calculation} formula
\item  \textbf{Dynamic} consideration of  user actions (e.g. starting/stopping of a VM) that may influence SLA applicability
\item  Downtime due to\textbf{ maintenance}
\end{enumerate}

\item  General assurance mechanisms in terms of faults that may be accredited to
$3^{rd}$ party services/equipment

\begin{enumerate}
\item  For example testing of general Internet connectivity on the client side
\item  Front-end API availability of provider 
\item  Monitoring daemon not running for an interval at the client side. This can be covered by appropriate assumptions, e.g. if logs are not available for an interval then the services are considered as available
\end{enumerate}

\item  Logging mechanisms that persist the necessary range and type of information
\end{enumerate}

\paragraph{System Setup}

System setup should include a running Cloud service. Furthermore, it should
include the benchmark/auditing code (according to the aforementioned
requirements taken under consideration) that is typically running externally to
the Cloud infrastructure, in order to cover the case that connectivity exists
internally in the provider but not towards the outside world.

\paragraph{Workload}

Given that this is a daemon-like benchmark, the concept of workload is not
directly applicable. The only aspect of workload that would apply would be for
the number of Cloud service instances to be monitored and the only constraint is
that these cover the preconditions of the SLA. However an interesting
consideration in this case may be the differentiation based on the complexity of
the observed service (in terms of number of virtual resources used), given that
this would influence the probabilities of not having a violation.

If we consider the case of a typical Cloud deployment at the IaaS level, we may
use N availability zones (AZ), in which M virtual machines are deployed. An
availability zone is typically defined as a part of the data center that shares
the same networking and/or power supply. Thus the usage of multiple AZs
eliminates the existence of a single point of failure for the aforementioned
risks. In the generic case, M is a 1xN vector, containing the number of VMs
deployed in each AZ. For simplicity purposes we assume that M is the same across
all AZs. If $P_{POW}$is the probability of power supply failure and $P_{AZNET
}$the probability of network failure, $P_{PH}$ the risk of the physical host in
which a VM is running to fail and $P_{VM}$ the risk of the VM to fail then,
assuming that these probabilities are mutually exclusive, depending on different
factors, the overall probability of failure for a deployment in one AZ is given
by Equation 24. The service is deemed as unavailable in one specific AZ if power
or network connectivity is lost across the AZ or if all VMs in that AZ are at the same time unavailable.

\begin{equation}
\begin{array}{l} {P_{OTH} =P_{POW} +P_{AZNET} } \\ {P_{NODE} =P_{PH} +P_{VM} }
\\ {P_{AZ} =P_{OTH} +\prod _{i=1}^{M}P_{NODEi}  =P_{OTH} +P_{NODE} ^{M} } \end{array}\end{equation}
\begin{center}Overall probability for an AZ to fail\end{center}

If M VMs are deployed in each one of the N AZs, then the overall
failure probability, assuming that the various AZs have similar power and
network probability failures and given that we are not aware of the affinity of
VM placement across physical nodes thus we can assume that different physical
hosts are used for each VM, we have the overall failure (and overall
unavailability) probability to be given by Equation 25.
\begin{equation}P_{OVERALL} =(P_{OTH} +P_{NODE} ^{M} )^{N} =\sum _{k=0}^{N}\frac{N!}{K!(N-K)!} P_{OTH} ^{K} P_{NODE} ^{M(N-K)}  \end{equation}
\begin{center}Overall service unavailability
probability\end{center}

Thus the significant factors that indicate the complexity (M and N) can be used
as a generic metric of ``workload''. Furthermore they can be used to classify
results of Equation 24 in categories according to the service complexity.

\subsection{Discussion}

While availability has been defined in the literature in various ways, existing
mainstream public Clouds such as Amazon, Google and Azure have separate
definitions, which may be similar but not identical even to each other. Thus
direct comparison of providers based on these metrics can not be achieved and
especially benchmarked against the guarantees they advertise in their
respective SLAs. In this section, an analysis is performed regarding the
similarities of provider definitions and how these can lead to guidelines
regarding the implementation of benchmarking clients for identifying provider
fulfillment of the issued Service Level Agreements towards their customers.

Furthermore, we define a simple yet directly comparable (between providers)
metric of SLA adherence. Classes of workload can be identified based on the
size and deployment characteristics of the measured service thus further
refining the aforementioned comparison.
\section{Operational Risk}\label{sec:Risk}

\subsection{Goal and Relevance}

Cloud services are commonly inscribed with performance guarantees which play important role on cloud resource-management.
These performance guarantees are actually levels of performance described in Service-Level-Agreements~(SLA) having been agreed upon between cloud providers and customers. The violation of SLAs impose financial penalties on cloud providers as well as reputation cost in cloud market.  Therefore, a metric providing the assessment of service performance-levels and of possible SLA violations is needed. Such performance levels can be characterized as expected levels for systems in order to perform without implications. Another definition of the expected system performance-levels may be the performance of common systems whose performance functions as baseline factor to assess other systems. Considering the performance expectations of systems, the \textit{operational risk} is defined as the likelihood of service performance to be on expected levels.

The operational risk metric complements the purpose of the other metrics proposed in previous sections. Risk management in cloud computing reflects the necessity of knowing the severity of changes in cloud infrastructures by evaluating the implications.
While metrics specialized for particular cloud features~(e.g., Elasticity, Isolation and Availability) measure the respective system performance regarding the features, the role of operational risk is different. The operational risk can be measured for any of the mentioned features by defining their type of performance and their respective expected levels. In other words, an operational risk metric can reflect the risk that the system performance deviates from the expected performance levels despite the type of performance.  

Having a representative metric of operational risk contributes to a higher level approach towards the performance evaluation of complex systems as the cloud ecosystems. In addition, the metric provides a level of abstraction of the overall system performance to stakeholders who concern about metrics depicting levels of system risk.

\subsection{Prerequisites and Context}

In this section, we describe the operational risk of cloud services deriving from their performance levels when running in cloud infrastructures. The term $risk$ is also used in literature for describing security issues of cloud but the notion of security is out of scope of the presented metric.

The \textit{service performance} in clouds refers to multiple kinds of performance
that can be measured in cloud services. A possible distinction among these
performance types is the level of service (e.g., IaaS, PaaS, SaaS) in cloud system.
The service level indicates the service performance for that level and also the suitable metrics measuring the performance . For example, the \textit{response-time} is useful for measuring service performance in PaaS, SaaS levels while the performance on leveraging service resources (e.g., CPU, memory) is addressed by IaaS-oriented metrics. In this section, we focus on the service performance in the level of IaaS, thus the operational risk utilizes IaaS-level metrics and refers to performance levels which reflect the resource utilization by cloud services.

\subsection{Proposed Operational Risk Metrics}

The idea behind the \textit{operational risk} metric proposed in this section is to quantify the variation of service performance between the service running in a dedicated environment and
the performance when the service running in another, non-dedicated environment. The
former provides an isolated environment where resources are always available to the service under test. In contrast, due to concurrent operations of services in a non-dedicated
environment~(i.e., common cloud system), services share the resources resulting in performance interference among the services~\cite{Zhuravlev2010}. This interference 
affects the performance delivered in services and in turn, this performance may deviate from the service performance in the dedicated environments. This deviation directly impacts the service performance resulting in its degradation.

For this reason, the term $risk$ is used to imply the likelihood that the actual service performance in the non-dedicated cloud will deviate from the demanded performance a service requires, which is reflected by the performance in the dedicated system.

\subsubsection{Related Metrics} 
\begin{figure}[htb]
\centering
\includegraphics[width=0.8\columnwidth]{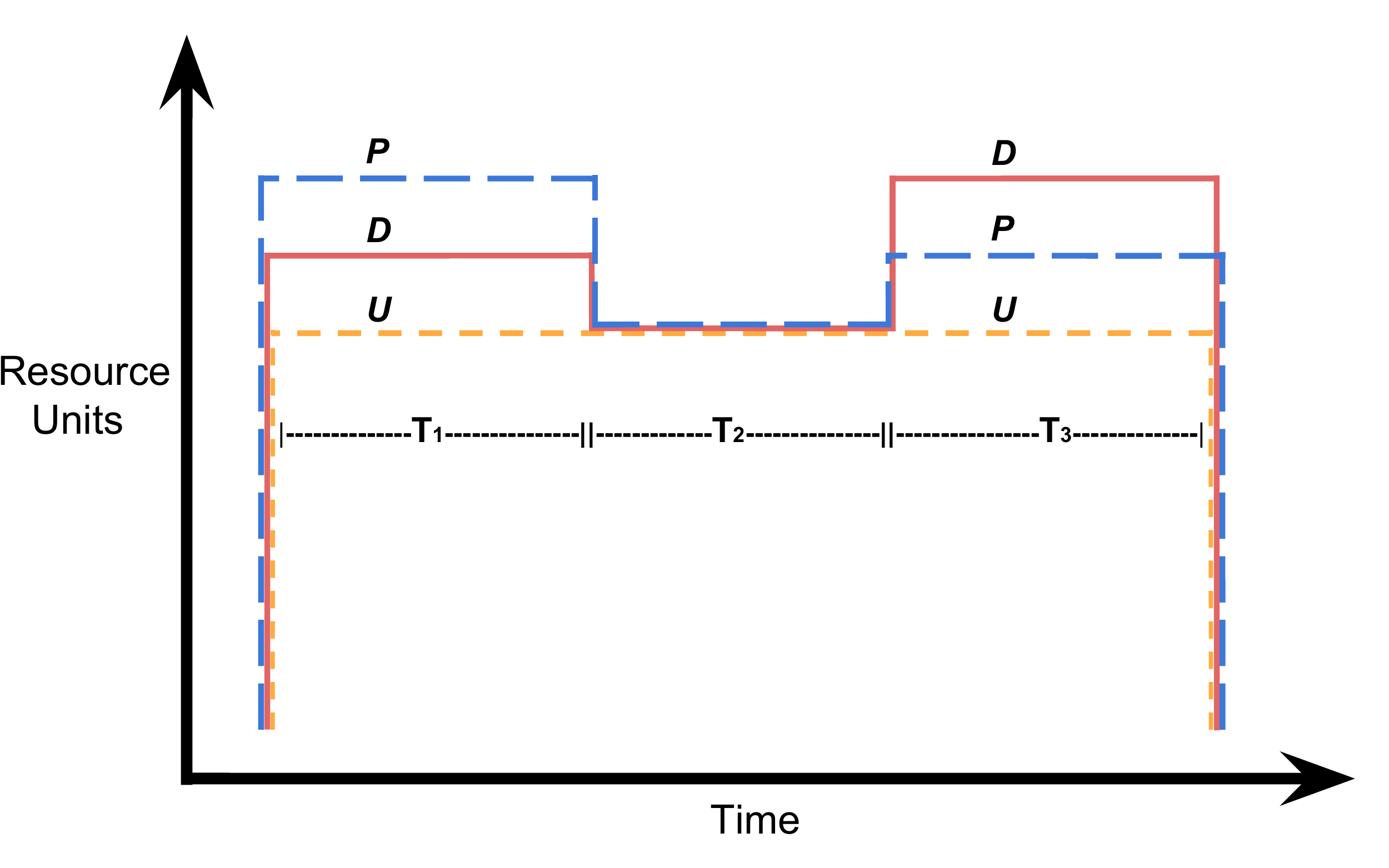}
\caption{Metrics composing the Operational Risk metric for cloud service.}
\label{fig:resmetrics}
\end{figure}

The risk metric reveals an additional aspect of the service performance in the cloud which is the degree to which service performance is degraded. Figure~\ref{fig:resmetrics} shows the three metrics referring to resource type of a cloud service. The current usage~($U$) of the resource, the demanded~($D$) amount the service requires and the provisioned~($P$) amount of resource which is the upper limit of resource the service is able to consume and is given by the resource manager. 

As Figure~\ref{fig:resmetrics} shows, $D$ line can be higher than the $U$ line as a service may not receive what it requires due to contention issues in the system, e.g., resource overcommitment. We consider that usage $U$ cannot be higher than the demanded amount $D$~($U\rlap{\kern.50em$|$}> D$) because the real usage of resource cannot exceed the requirements of service in resources. Similarly, $U\rlap{\kern.45em$|$}> P$ as the actual resource consumption can only reach up to the resource limit that has been set by the resource manager.

\paragraph{Relative Levels of Metrics}

Considering the three mentioned metrics, we show the possible cases about the relative levels of the metrics. In an ideally auto-scaling and isolated environment, the service is provisioned and consumes the demanded amount of resource~($P=D=U$), as in period $T_{2}$ in Figure~\ref{fig:resmetrics}. The resource manager provides the demanded resources~($P=D$) and the service utilizes all the demanded resources~($D=U$) without interfering with other co-hosted services.

However, the case of the three coinciding lines is not applicable in real environments. An auto-scaling mechanism is not perfectly accurate and creates incompatibilities between the provisioned and the demanded resources. When a service demands fewer resources than it has been provisioned~($D<P$), then the system hosts an over-provisioned service. In the opposite way, the system has under-provisioned resources~($D>P$) to service. The two mentioned cases are depicted in periods $T_{1},T_{3}$ in Figure~\ref{fig:resmetrics} respectively.

Another possible case of the metric levels is the relative position between $D$ and $U$. As mentioned before, the real usage cannot be higher than the demanded usage, however coinciding lines of $U$ and $D$ is not always the case due to contention issues in the system. We call the deviation between $D$ and $U$, which is always non-negative~($D\geq U$), as \textit{contention-metric} and it presents the gap between the demanded amount of resource and the real consumption of resource at specific time. The higher the value of the contention-metric, the more severe is the resource-contention which is experienced by the service; hence degrading the service performance. For instance, in Figure~\ref{fig:resmetrics}, the service experiences more contention in its resources during period $T_{3}$  than during period $T_{1}$ as the gap between the resource lines $D,U$ is bigger in the former period.

\subsubsection{Definition of Metric}

The operational risk metric presented in this Section incorporates the three resource metrics mentioned before and splits the metric functionality into two partial metrics, the \textit{provision risk}~($r_{p}$) and the \textit{contention risk}~($r_{c}$). The objective of the two partial metrics is the combination of two cloud features, elasticity and isolation. This aggregation will benefit the metric users to have more comprehensive view of the system performance and adjust performance issues towards their needs.

\paragraph{Provision Risk}

We define as risk of provisioned resources or \textit{provision risk}~($r_{p}$), the degree to which the demanded resources meet the provisioned resources.

\begin{equation} \label{eq:rp}
r_{p} =\frac{1}{T} \int\limits_{T} \frac{P_{t}-D_{t}}{P_{t}} \textit{d} t ,\qquad             [-1,1]
\end{equation}

where $P_{t}$ and $D_{t}$ are the provisioned and the demanded resources respectively at time $t$. The $T$ value is the time period which we measure the two metrics. The integral value of $r_{p}$
measures the relative squashed area enclosed by the resource values of $P$ and $D$ for time period $T$. The value of $r_{p}$ ranges between $[-1,1]$ indicating an under-provisioning situation when $r_{p} \in [-1,0)$ and an over-provisioning case when $r_{p} \in (0,1]$. The zero value indicates an accurate provision of resources according to the demanded resources.
Thus, the closer the value of $r_{p}$ is to zero, the less risk is indicated for the service.

\paragraph{Contention Risk}

Similarly, we define the \textit{contention risk}~($r_{c}$). This metric utilizes the resource values of demanded and used resources over time $T$. The metrics $D_{t}$ and $U_{t}$ indicate the values of the respective metrics at time $t$.

\begin{equation} \label{eq:rc}
r_{c} = \frac{1}{T} \int\limits_{T} \frac{D_{t}-U_{t}}{D_{t}} \textit{d} t,\qquad [0,1]
\end{equation}

The $r_{c}$ value is non-negative as the amount of used resources cannot exceed the amount of demanded resources. The higher the value of $r_{c}$, the more risk is estimated for a service to not receive the demanded performance due to resource contention.

\paragraph{Service Risk}

The risk for a cloud service should be composed by the combination of the two aforementioned risk metrics in order to have better overview on the service status regarding risk levels.

We define the \textit{risk of service}~($r_{e}$) as the degree to which a cloud service performs as expected. The expected performance derives from the expected performance of an elastic system and the expected performance with respect to the experienced contention in a service. Both performance levels contribute to evaluate if a system supplies the service adequately with enough resources.

\begin{equation}\label{eq:re}
r_{e} =  w_{p} \times |r_{p}| + w_{c} \times r_{c} = \frac{1}{T} \left(w_{p} \times  \int\limits_{T} \frac{|P_{t}-D_{t}|}{P_{t}} \textit{d} t +
w_{c} \times \int\limits_{T} \frac{D_{t}-U_{t}}{D_{t}} \textit{d} t  \right)
 , \qquad [0, 1]
\end{equation}
$\text{with}\qquad w_{p},w_{c} \in [0,1], w_{p}+w_{c} =1$.
\\

For the metric $r_{e}$, the difference between $P_{t}$ and $D_{t}$ is an absolute value because we do not focus on the provisioning type of risk but only for the level of severity that the difference between the two metrics reflects. Factors $w_{p},w_{c} \in [0,1]$ are used to weight the operational risk value according to user needs.

\paragraph{System Risk}

The operational risk of cloud system should be an overview of the corresponding risk values of the services in the system. The aggregation method of service risks, which calculates the system-risk value, deviates according to user needs. The method will probably be a variability metric which combines values according to different purposes. 
For example, cloud providers who want to test the risk levels of the system may use quantile values to depict the risk level of some service groups.
For cloud customers, who want to deploy cloud services into an environment with stable risk levels, a variability metric like the interquartile range~($IQR$) will show the dispersion of service risks in the system and thus, the performance variability of the system.

\subsection{Measurement Methodology}

To measure effectively the operational risk of cloud systems, we have to determine which monitored data are the respective metrics that operational risk is built upon. Additionally, the resource type has to be decided in order the appropriate resource metrics to be declared. We focus on the most common resource types that are currently available to clouds, i.e., CPU, memory, network and storage.  
\paragraph{Metrics $P$ \& $U$}The metric definitions of provision~($P$) and usage~($U$) are straight-forward and one can easily monitor the respective values. Metric $P$ refers to the capacity that resource manager has provisioned to service and metric $U$ is the actual capacity that service uses. In any resource type, the corresponding metrics of $P$ and $U$ can be readily monitored. 

\paragraph{Metric $D$}
Although it may be confusing how the demanded amount of the aforementioned resource types is estimated, there is currently enough research on that topic. For the resource type of CPU, prediction models have been introduced in ~\cite{Isci2010} and contemporary cloud monitors provide the metric of $CPU\_Demand$. For the memory resource, the demanded amount can also be estimated as in~\cite{Zhou2004} and used as the possible memory capacity that service needs at specific time. For the resource types of network and storage, the metric $D$ is simpler to be calculated as the demanded capacities are either the size or the number of requests of that resource received by the resource manager. The resource manager, after collecting the requests, handles a subset of these requests~(i.e., metric $U$) due to the system load.

\paragraph{Weights $w_{p},w_{c}$}
The values of the two weights $w_{p},w_{c}$ represent the importance of the two metrics $r_{p},r_{c}$ respectively. One has to consider the purpose of benchmarking the cloud system in order to define similarly the weights. The \textit{contention risk} may be more important for testing the impact of co-location of cloud services~($w_{p} < w_{c}$) whereas the \textit{provision risk} represents better the operational risk when selecting the most promising elastic policy for a system.

\subsection{Related Work}

Risk management and resource contention are not new subjects in cloud research. 
Risk management has been defined as decision paradigm in Grids~\cite{Djemame2006} where selection among multiple infrastructures should have been taken into account to host an application. In this work and also in the extended work of~\cite{Ferrer2012}, historical records about SLA violations are used to assess the system risk for an incoming application to fulfill the agreed objectives. The risk levels are also evaluated according to the provisioned mismatches of the elasticity mechanism. Our approach of operational risk considers the cloud feature of elasticity and incorporates it with the risk of performance interference of services.

In~\cite{Tang2011}, the authors investigate the interference of services in memory. The presented results on performance degradation use as baseline-level the performance of a service running alone in the system under test. In~\cite{Zhuravlev2010}, a similar approach measuring the performance degradation is used with a more explicit definition of the degradation. The authors define the relative performance degradation of memory components according to the performance level of a running-alone service. We extend this related work considering the impact of elasticity in the system. The degradation of performance due to resource contention is affected by the elastic mechanisms in clouds and this emerges the necessity of incorporating the two notions.

\subsection{Discussion}

In this section, we present the feature of operational risk in clouds and introduce a representative metric evaluating the risk in cloud services and systems not to perform as expected. 

The notion of risk differentiates from the elasticity feature in cloud because elasticity takes into account the provisioned and the demanded values of service resource to cope with the service load. 
In contrast, both performance isolation and availability utilize demand and usage metrics for different reasons.
Performance isolation concerns about the contention that a service may experience and evaluates the severity of the interference among services while availability evaluates the periods where demand is present but the usage of resources cannot be achieved.
 
The operational risk incorporates the three resource-level metrics and complements the other metrics described in previous sections in order to assess the severity of performance degradation regarding the mentioned cloud features.

\subsubsection{Usability}

The operational risk metric can be used as evaluation of cloud services and systems to assess whether the performance guarantees are met.

The measurement of service-performance in the cloud is important for both customer and cloud provider. The customer wants to maximize profit by delivering good QoS to clients. Therefore, the service performance is
utilized by the customer to check the progress of service-runtime as well as to compare and select the cloud environment which meets the service needs the most.

For cloud providers, they are interested in the levels of service performance because they are burdened with financial fees when SLA violations occur. Moreover, there is a reputation cost for cloud providers when not delivering good performance results to customers, thus, providers also utilize service-performance metrics to maintain competition in the cloud market.

\section{Conclusion}

Because cloud computing services, and the systems underlying them, already account for a large fraction of the information and communication technology (ICT) market, 
understanding their non-functional properties is increasingly important.
Building for this tasks cloud benchmarks could lead to better system design, tuning, and selection. Responding to this need, in this report, we highlight the relevance of new non-functional system properties emerging in the new context of cloud computing, namely elasticity, performance isolation, availability and operational risk. We discuss these properties in depth and select existing or propose new metrics that are capable to quantify these properties. Thus, for these four properties we lay a foundation for benchmarking cloud computing settings.

As future activities, we plan to conduct real-world experiments that underline the applicability and usefulness of the proposed metrics, also refining the corresponding measurement approaches. As a next step, we are working on an extensive review of existing cloud-relevant benchmarks and connected domains like big data, web services, and graph processing.

\cleardoublepage
\pagenumbering{gobble}
\pagestyle{empty}
\renewcommand\bibname{References}

\bibliographystyle{abbrnat2}
\bibliography{maindoc}

\end{document}